\documentclass[11pt,letterpaper]{article}
\usepackage{pslatex}
\usepackage{apacite}
\usepackage{graphicx}
\usepackage{textcomp}
\usepackage[all]{nowidow}
\usepackage{amsfonts}
\usepackage[font=small,skip=0pt]{caption}
\usepackage{mathtools}
\usepackage{float}
\usepackage{courier}
\usepackage{placeins}
\usepackage{sidecap}
\usepackage{lineno}
\usepackage{authblk}
\usepackage{caption,setspace}
\usepackage{xcolor}
\usepackage{changepage}
\usepackage{subcaption}
\usepackage{booktabs}
\usepackage{rotating}
\usepackage{color}
\definecolor{Blue}{RGB}{50,50,200}
\definecolor{Red}{RGB}{200,50,50}

\usepackage[letterpaper, margin=0.9in]{geometry}
\usepackage{pdflscape}
\usepackage{url}

\title{Pragmatic inference and visual abstraction enable contextual flexibility during visual communication}
\date{}

\author[a,c]{Judith E. Fan}
\author[a]{Robert X.D. Hawkins}
\author[b]{Mike Wu}
\author[a,b]{Noah D. Goodman}

\affil[a]{Department of Psychology, Stanford University}
\affil[b]{Department of Computer Science, Stanford University}
\affil[c]{Department of Psychology, University of California, San Diego}

\begin{document}
\maketitle

\begin{abstract}
\hyphenpenalty=1000 

Visual modes of communication are ubiquitous in modern life --- from maps to data plots to political cartoons. Here we investigate drawing, the most basic form of visual communication. Participants were paired in an online environment to play a drawing-based reference game. On each trial, both participants were shown the same four objects, but in different locations. The sketcher's goal was to draw one of these objects so that the viewer could select it from the array. On `close' trials, objects belonged to the same basic-level category, whereas on `far' trials objects belonged to different categories. We found that people exploited shared information to efficiently communicate about the target object: on far trials, sketchers achieved high recognition accuracy while applying fewer strokes, using less ink, and spending less time on their drawings than on close trials. We hypothesized that humans succeed in this task by recruiting two core faculties: visual abstraction, the ability to perceive the correspondence between an object and a drawing of it; and pragmatic inference, the ability to judge what information would help a viewer distinguish the target from distractors. To evaluate this hypothesis, we developed a computational model of the sketcher that embodied both faculties, instantiated as a deep convolutional neural network nested within a probabilistic program. We found that this model fit human data well and outperformed lesioned variants. Together, this work provides the first algorithmically explicit theory of how visual perception and social cognition jointly support contextual flexibility in visual communication.

\end{abstract}

\newpage

\section*{Introduction}

From ancient etchings on cave walls to modern digital displays, the ability to externalize our thoughts in visual form lies at the heart of key human innovations (e.g., painting, cartography, data visualization), and forms the foundation for the cultural transmission of knowledge \cite{tomasello2009cultural,donald1991origins}. 
Perhaps the most basic and versatile visualization technique is drawing, the earliest examples of which date to at least 40,000 years ago \cite{hoffmann2018u,Aubert:2014jy}, and which can yield images ranging from photorealistic renderings to schematic diagrams. 
Even in the simple case of sketching an object in the world, there are countless ways of depicting that object. 
How do drawings, despite spanning such a broad range of appearances, reliably convey meaning? 

On the one hand, recent work in computational vision suggests that the identity of an object depicted in a drawing can be derived from its visual properties alone \cite{FanCommon2018}.
These results are consistent with evidence from other domains, including developmental, cross-cultural, and comparative studies of drawing perception. 
For example, human infants \cite{hochberg1962pictorial}, people living in remote regions without pictorial art traditions and without substantial contact with Western visual media \cite{kennedy1975outline}, and higher non-human primates \cite{tanaka2007recognition} are able to recognize line drawings of familiar objects, even without prior experience with drawings.
Together, these findings suggest that the ability to perceive the correspondence between drawings and real-world objects arises from a general-purpose neural architecture evolved to handle variation in natural visual inputs \cite{Sayim:2011bz,gibson2014ecological}. 

On the other hand, influential work in philosophy has emphasized the role of cultural and social context in determining how drawings denote objects \cite{goodman1976languages}.
This perspective is consistent with substantial variation in pictorial art traditions across cultures \cite{gombrich1989story,gombrich1969art} and the existence of culturally-specific conventions for encoding meaning in pictorial form \cite{boltz1994origin,allen2000middle}. 
Further support for the importance of social context has also come from recent laboratory studies of visual communication, which have found that pairs of interacting participants can produce drawings that are referentially meaningful to their partner in context, even when these drawings do not strongly resemble any particular real-world referent out of context \cite{Garrod:2007wk,fay2010interactive,Galantucci:2005uh}. 

Towards reconciling these perspectives, the current paper explores the hypothesis that visual information and social context jointly determine how drawings convey meaning.
To evaluate this hypothesis, we investigated how the drawings people produce varied across communicative contexts, and found that people adapted their drawings accordingly, producing detailed drawings when necessary, but simpler drawings when sufficient.
To explain these findings, we developed a computational model of visual communication that embodied two core faculties: visual abstraction, the capacity to judge the correspondence between a real-world object and a drawing of it; and pragmatic inference, the ability to judge what information is not only \textit{valid} to include in a drawing, but also \textit{relevant} in context  \cite{goodman2016pragmatic,grice1975syntax,abell2009canny}.
This model was instantiated as a deep convolutional neural network visual encoder nested within a probabilistic program that inferred which drawings would be most informative in context.
We found that our full model fit the data well and outperformed lesioned variants, providing a first algorithmically explicit theory of how visual perception and social cognition jointly support contextual flexibility in visual communication.

\section*{Results}

\subsection*{Effect of context manipulation on communication task performance}



To investigate visual communication in a naturalistic yet controlled setting, we employ a drawing-based reference game paradigm.
This reference game involves two players: a \textit{sketcher} who aims to help a \textit{viewer} pick out a target object from an array of distractor objects by representing it in a sketch. 
Such games, which have long provided a source for intuitions in the philosophy of language \cite{wittgenstein1953philosophical,Lewis69_Convention}, have also proven to be a valuable experimental tool for systematically eliciting pragmatic inferences about language use in context \cite{goodman2016pragmatic,kao2014formalizing,goodman2013knowledge,frank2012predicting}, especially the ability of speakers to compose utterances that are informative \cite{grice1975syntax,wilson1986relevance} yet parsimonious \cite{zipf1936psycho} during verbal communication. 
Here we generalize this methodology to understand the role of pragmatic inference during visual commmunication. 

\begin{figure}[htbp]
\centering
\includegraphics[width=0.42\textwidth]{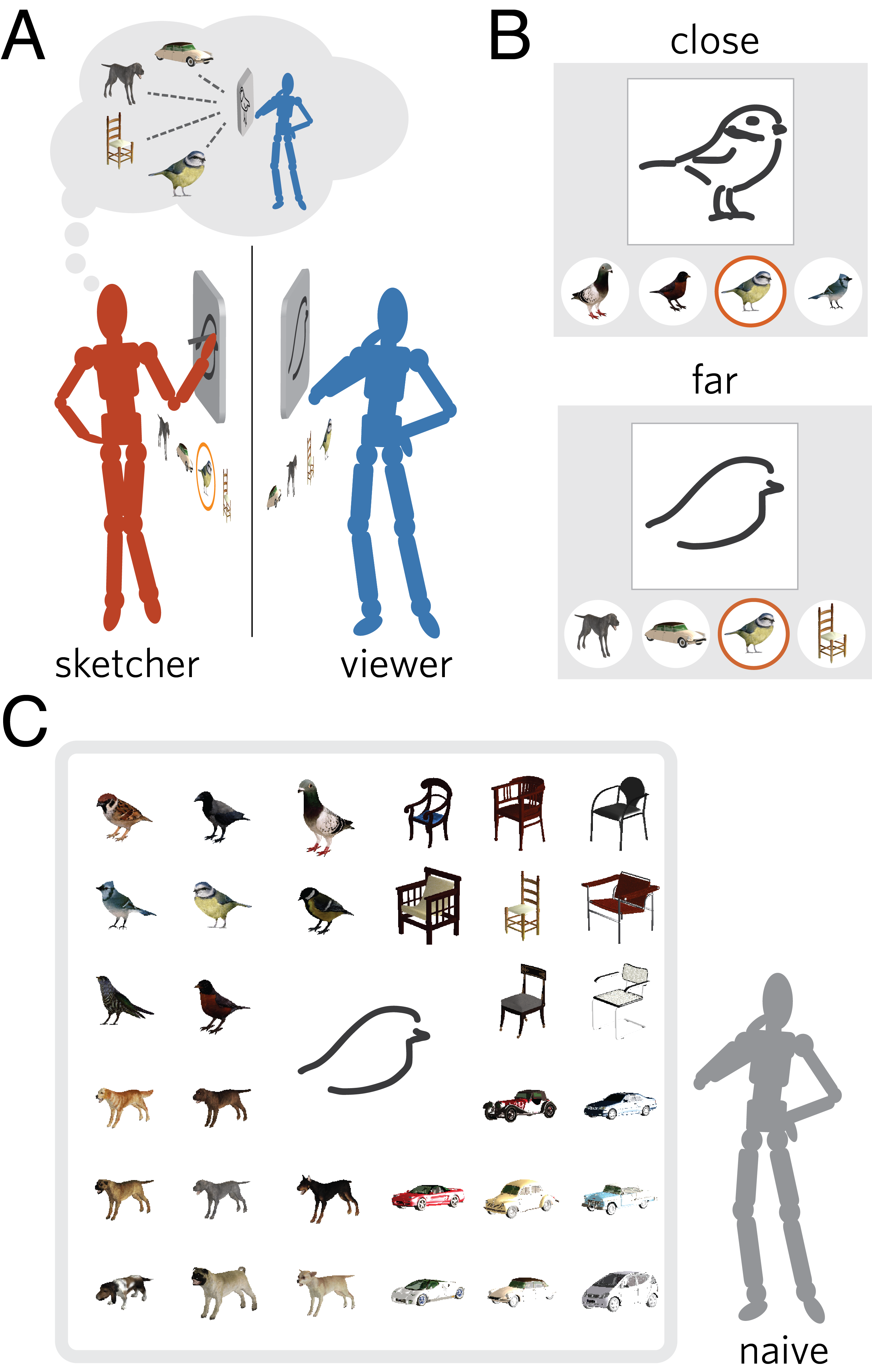}
\caption{(A) Communication task. Participants were paired in an online environment to play a drawing-based reference game and assigned the roles of sketcher and viewer. On each trial, the sketcher's goal was to draw one of these objects so that the viewer could distinguish the target from three distractor objects. (B) Context manipulation. Distractor similarity to target was manipulated across two context conditions: in close contexts, the target and distractors belonged to the same basic-level category, while in far contexts, the target and distractors all belonged to different basic-level categories. (C) Recognition task. Naive participants were presented with a randomly sampled sketch from the communication experiment and an array containing all 32 objects used in the experiment, and were instructed to identify the best-matching object.}
\label{task_display}
\end{figure}
\begin{figure*}[htbp]
\centering
\includegraphics[width=0.95\textwidth]{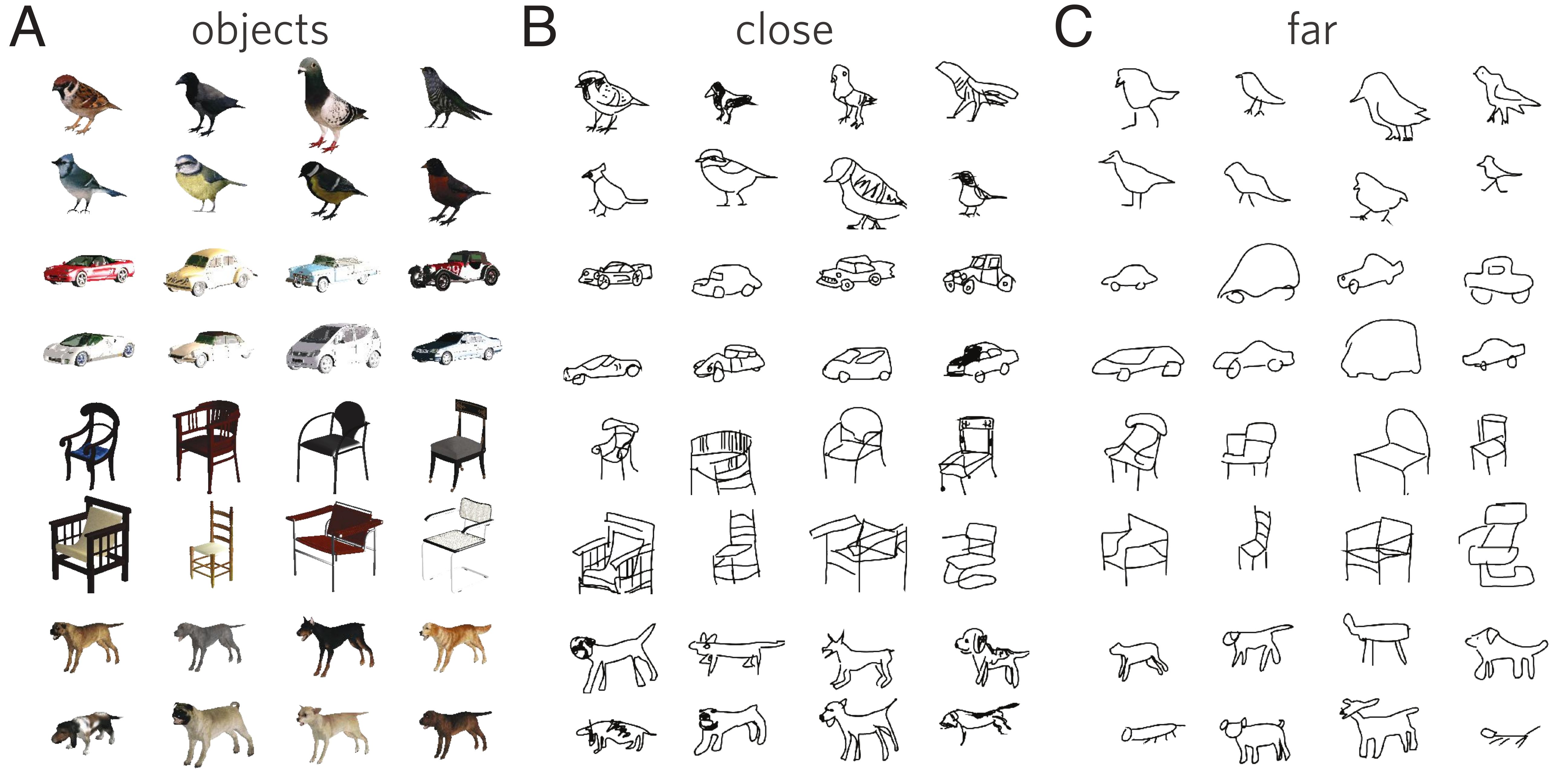}
\caption{(A) Object stimuli. (B) Example sketches produced in close context condition. (C) Example sketches produced in far context condition.}
\label{sketch_gallery}
\end{figure*}

In our experiment, participants (N=192) were paired in an online environment and communicated with their partner only via a drawing canvas (Fig.~\ref{task_display}A). 
Each trial, both participants were shown a set of four real-world objects, but object locations were randomized for each participant so that they could not use object location information to solve the task. 
The sketcher's goal on each trial was to draw one of these objects --- the target --- so that the viewer could pick it out from the array. 
There were 32 objects in total belonging to four basic-level categories (i.e., bird, car, chair, dog), that were rendered in the same three-quarter pose, under identical illumatination, and on a gray background, so participants could not use pose, illumination, or background information to distinguish them. 
Each object was randomly assigned to exactly one set of four objects, and each set of four objects was presented four times each, such that each object served as the target exactly once. 
Across trials, the similarity of the distractors to the target was manipulated, yielding two types of communicative context that appeared in a randomly interleaved order: close contexts, where the target and distractors all belonged to the same basic-level category, and far contexts, where the target and distractors belonged to different basic-level categories (Fig.~\ref{task_display}B). 
We predicted that while sketchers would be generally successful at conveying the identity of the target, their sketching behavior would systematically differ between the two contexts. 
Specifically, we predicted that sketchers would invest more time and ink in producing their sketches in close contexts, but still produce sufficiently informative sketches with less time and ink in far contexts (Fig.~\ref{sketch_gallery}). 

\begin{figure*}[htbp]
\centering
\includegraphics[width=0.9\textwidth]{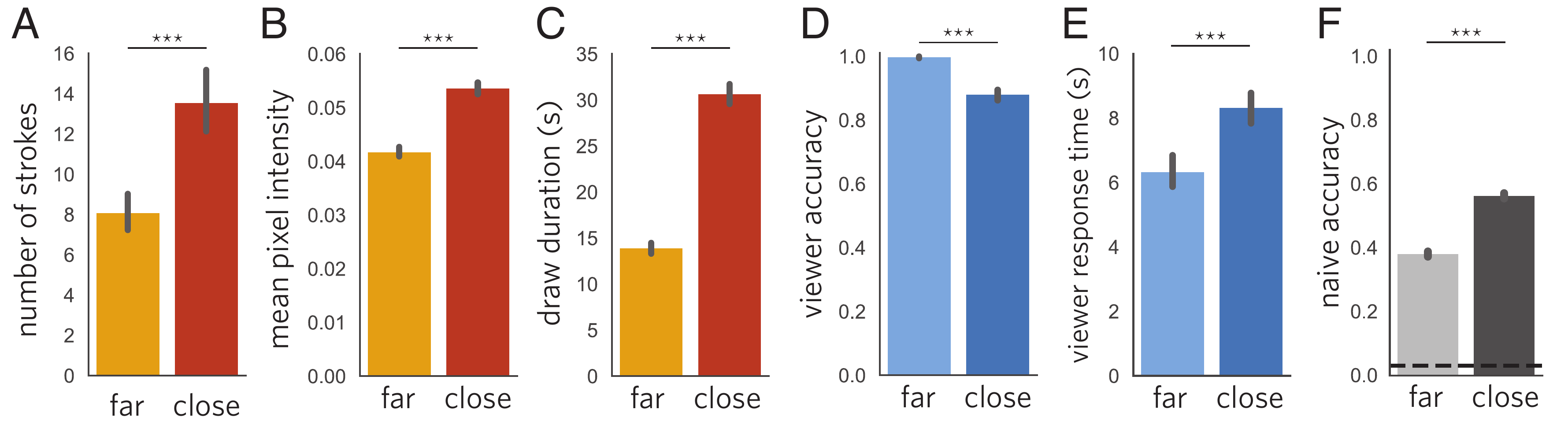}
\caption{(A-C) Mean number of strokes, amount of ink, and time spent producing sketches in each context condition. (D-E) Target identification performance in context during communication task. (F) Target identification performance out of context during recognition task.}
\label{task_performance}
\end{figure*}

Consistent with our prediction, we found that viewers were highly accurate overall at identifying the target from the sketches produced (proportion correct: 93.8\%, 95\% CI: [92.7\%, 94.8\%], estimated by bootstrap resampling participants). 
Moreover, we found that sketchers spent less time (close: 30.3s, far: 13.7s, \textit{p}$<$0.001), applied fewer strokes (close: 8.03 vs. far: 13.5, 95\% CI of difference: [3.75, 7.90], \textit{p}$<$0.001), and used less ink (proportion of canvas filled; close: 0.054, far: 0.042, 95\% CI of difference: [0.01, 0.014], \textit{p}$<$0.001) to produce their sketches in the far condition than in the close condition (Fig.~\ref{task_performance}A-C). 
Despite the relative sparsity of sketches in the far condition, viewers were near ceiling at identifying the target on these trials (far: 99.7\%, 95\% CI: [0.993, 0.999]; close: 87.9\%, 95\% CI: [0.858, 0.899], Fig.~\ref{task_performance}D), and took less time to make these decisions than on close trials (far: 6.32s vs. close: 8.32s, 95\% CI of difference: [-2.748, -1.251], Fig.~\ref{task_performance}E).

\subsection*{Effect of context manipulation on sketch recognizability}

A natural explanation for these findings is that the two context conditions differed in how much information was required to identify the target. 
Specifically, sketchers invested greater time and ink in close contexts to strengthen the correspondence between their sketch and the target object, out of necessity, while they could still succeed in far contexts with sketches that were less costly to produce.
To evaluate this possibility, we recruited another group of naive participants (N=112) to perform a sketch recognition task that yielded estimates of how strongly each sketch corresponded to every object in the communication experiment. 
On each trial of this recognition experiment, participants were presented with a sketch and an array containing all 32 objects, and were instructed to identify the object that best matched each sketch from the array (Fig.~\ref{task_display}C). 
Across trials, sketches were randomly sampled from the original communication experiment such that no two sketches produced by the same participant appeared in a single recognition experimental session. 
Consistent with our hypothesis, we found that close sketches were matched with their corresponding target object more consistently than far sketches were (close: 54.2\%; far: 37.5\%; $Z$=14.1, $p <$0.001), although sketches from both context conditions were successfully matched at rates greatly exceeding chance ($p$s $<$ 0.001).







\subsection*{Computational model of contextual flexibility in visual communication}

Our empirical findings suggest that sketchers spontaneously modulate the amount of information they convey about the target object according to the communicative context. 
Such contextual flexibility argues against the notion that visual communication is constrained exclusively by the appearance of the target object, but rather is systematically influenced by contextual information that is shared between the sketcher and viewer. 
Moreover, it suggests an analogy to how shared context influences what people choose to say during verbal communication, a key target of recent advances in computational models of pragmatic language use \cite{frank2012predicting,goodman2013knowledge,franke2016probabilistic,bergen2016pragmatic}.
Leveraging these advances, we propose that human sketchers determine what kind of sketch to produce in context by deploying two main faculties: \textit{visual abstraction}, which here refers to the ability to judge how well a sketch evokes a real object, and \textit{pragmatic inference}, which here refers to the ability to judge which sketches will be sufficiently detailed to be informative about the target object in context, but no more detailed than necessary. 
To test this proposal, we developed a computational model of the sketcher that embodies both visual abstraction and pragmatic inference, and was instantiated as a deep convolutional neural network nested within a probabilistic program. 
Constructing such a model allowed us to use formal model comparison to evaluate the contribution of each component for explaining our empirical findings, as well as make quantitative predictions about visual communication behavior in novel contexts.

\begin{figure*}[htbp]
\centering
\includegraphics[width=0.9\textwidth]{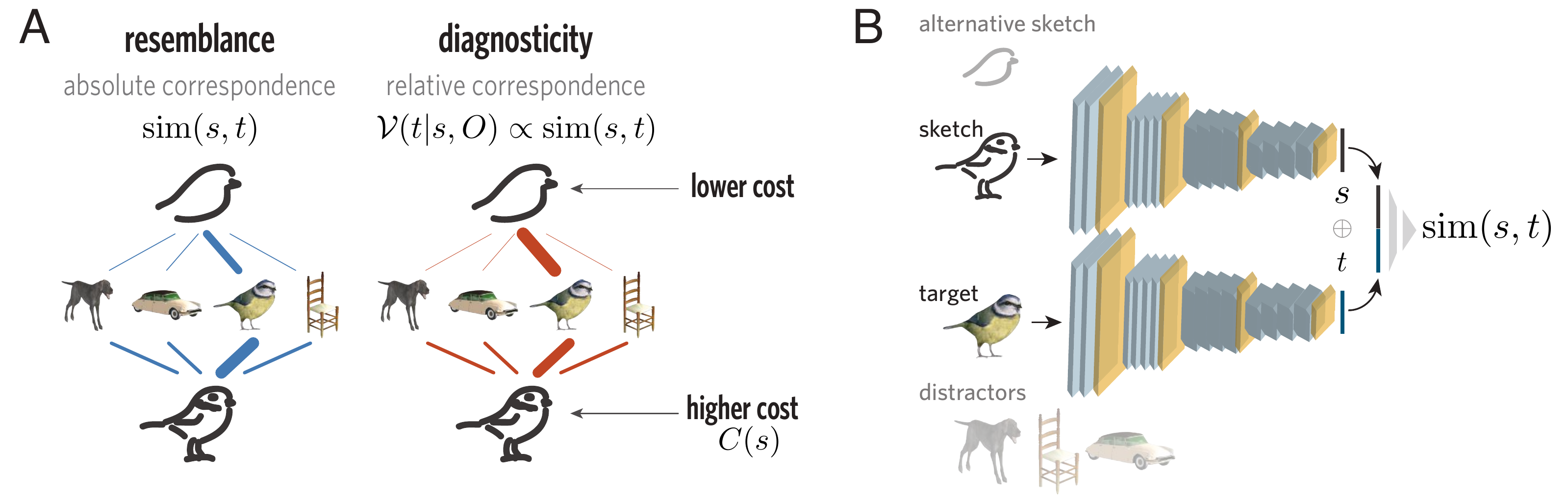}
\caption{
(A) Schematic containing an example context and two candidate sketches under consideration by the sketcher, $\mathcal{S}$. 
The thickness of each blue line reflects the absolute strength of the correspondence between a candidate sketch and object in context (resemblance). 
The thickness of each red line reflects the relative strength of the correspondence between a candidate sketch and each object, compared to its correspondence to the other objects in the context (diagnosticity). 
A sketch's informativity was hypothesized to depend on both its resemblance and diagnosticity. 
The sketcher expects the viewer, $\mathcal{V}$, given the sketch and context, to select the target object proportional to the strength of the correspondence between the sketch and target object. 
All else equal, the sketcher is assumed to prefer sketches that are less costly to produce.
(B) Architecture of the visual encoder used to predict the correspondence between sketches and objects, which consists of a base convolutional neural network and fully connected ``adaptor'' neural network. 
The parameters of the base neural network are trained on separate data and frozen, whereas the parameters of the adaptor neural network are trained on subsets of our experimental data. 
First, two identical branches of the base neural network are applied to a sketch and object to extract a feature vector for each image. 
Next, these feature vectors are concatenated and passed through the adaptor neural network to yield a sketch-object correspondence score.}
\label{model_schematic}
\end{figure*}

\subsubsection*{Defining communicative utility of sketches}

We define the sketcher, $\mathcal{S}$, to be a decision-theoretic agent that produces sketches, $s$, of the target proportional to their communicative utility, which is a function of a sketch and a context: $U(s,O)$.
In our experiment, a context is defined as:
$O = \{t,D\}$, where $t$ is the target object and $D$ is a set of three distractor objects, $D=\{d_1,d_2,d_3\}$.  
When deciding which sketch to produce, the utilities of each sketch are assumed to be normalized over the set of producible sketches via the softmax function: 
\begin{equation} \label{sketcher_distribution}
\mathcal{S}(s|O) = \frac {\exp [{U(s,O)]}} {\sum_{i} {\exp [U(s_i,O)]}}
\end{equation}
In principle, the space of all producible sketches is infinite and continuous, leading to an intractable sum.
In practice, we assume that the sketcher model chooses among a large but finite set of sketches: those actually produced by participants in our experiment.

We first introduce the utility function for our proposed pragmatic sketcher, $S_{prag}$, and then consider lesioned variants for comparison. 
Overall, this utility trades off between how informative a sketch is with how costly it was produce.
The sketcher judges a sketch's informativity to be a mixture of two quantities: one reflecting its absolute \textit{resemblance} to the target and the other its relative \textit{diagnosticity} in the presence of particular distractors (see Fig. \ref{model_schematic}B).  

Resemblance is defined by how strongly a sketch $s$ corresponds to the target $t$, i.e. $\textrm{sim}(s,t)$, which we estimate empirically using data from the recognition experiment (see Materials and Methods).
Diagnosticity is defined by the natural log probability that a simulated viewer agent, $\mathcal{V}$, would select the target object given the sketch and all objects in context, $\ln \mathcal{V}(t|s,O)$. 
The simulated viewer $\mathcal{V}$, in turn, is assumed to select the target object proportional to the correspondence between the sketch and the target, $\textrm{sim}(s,t)$, normalized by the sum of correspondences between the sketch and all four objects in context, again via the softmax function:
\begin{equation} \label{literal_viewer_score}
\mathcal{V}(t|s,O) \propto \frac {\exp\{\alpha \cdot \textrm{sim(s, t)}\}} {\sum_{i=1}^{4} \exp\{\textrm{sim}(s,o_i)\}}
\end{equation}
Here, $i$ indexes each object $o\in O$, and $\alpha$ is a scaling parameter determining the assumed optimality of the simulated viewer's decision policy. 
As $\alpha \rightarrow \infty$, the simulated viewer is more likely to choose the object with highest perceptual correspondence to the sketch. 
Intuitively, this means that the viewer is more likely to pick the correct object when the sketch corresponds more strongly to the target than to the distractors. 

To combine the resemblance and diagnosticity terms into a single informativity value, we introduce a weight parameter, $w_{d}$, that interpolates between them:
\begin{equation} \label{prag_interpolation}
I(s,O) = w_{d} \cdot \ln \mathcal{V}(t|s,O) + (1-w_{d}) \cdot \textrm{sim}(s,t). 
\end{equation} 
Combining these terms captures the intuition that a communicative sketcher seeks to produce a sketch that both resembles the target object and distinguishes the target from the distractors.
Finally, we define a sketch's cost, $C(s)$, to be a monotonic function of the amount of time taken to produce it, linearly transformed to lie in the range $[0,1]$. 
Putting these terms together, we have the full utility:
\begin{equation} \label{sketcher_utility}
U_{S_{prag}}(s,O) = w_i \cdot I(s,O) - w_c \cdot  C(s)
\end{equation}
where $w_i$ and $w_c$ are independent scaling parameters that are applied to the informativity and cost terms, respectively.
These parameters determine how strongly each term contributes to the overall utility of the sketch. 
This model contains four latent parameters: one each on informativity ($w_{i}$) and cost ($w_{c}$), one that balances between diagnosticity and resemblance within the informativity term ($w_{d}$), and one that tracks the optimality of the simulated viewer's decision policy ($\alpha$). 
We inferred these parameters from our data via Bayesian data analysis (see Materials and Methods).


\subsubsection*{Evaluating contribution of pragmatic inference}

We hypothesized that a pragmatic sketcher model that is sensitive to both context and cost would provide a strong fit to human sketch production behavior, as well as outperform lesioned alternatives lacking either component.
To test this hypothesis, we compare the full pragmatic model, ($S_{prag}$), with two nested variants with different utility functions:
a \textit{context-insensitive} sketcher, $S_{sim}$, in which the diagnosticity term is removed (i.e.,~$w_{d}{=}0$), leaving only the resemblance component in the informativity term; and  
a \textit{cost-insensitive} sketcher, $S_{nocost}$, in which the cost term is removed (i.e.,~$w_c=0$), leaving only the full informativity term. 


Our goal was to evaluate how well each model could produce informative sketches and appropriately modulate its behavior according to the context condition, and not necessarily to reproduce exactly the same sketch a particular participant had on a specific trial. 
As such, we aggregated all sketches of the same object produced in the same context condition, yielding 64 `prototype' sketches for each object-context category (e.g., basset sketch produced in a close context), which exhibited the average strength of object correspondence and cost in each category. 
Decisions by the sketcher model were generated at the same level of granularity, in the form of a probability distribution over these 64 prototype sketches. 
To generate these decisions, first we employed Bayesian data analysis to infer a posterior distribution over the four latent parameters in the model ($w_{i}$, $w_{c}$, $w_{d}$, $\alpha$). 
Next, we presented each model with exactly the same set of contexts that were presented to human sketchers in the communication experiment, and evaluated the posterior predictive probabilities that each model assigned to sketches in each object-context category, marginalizing over the posterior distribution over latent parameters. 
We conducted these inference and evaluation steps independently on five balanced splits of the dataset, providing an estimate of reliability and permitting side-by-side comparison with subsequent modeling results using the same splits for crossvalidation (see \textit{Evaluating contribution of visual abstraction}). 


\begin{figure*}[htbp]
\centering
\includegraphics[width=0.99\textwidth]{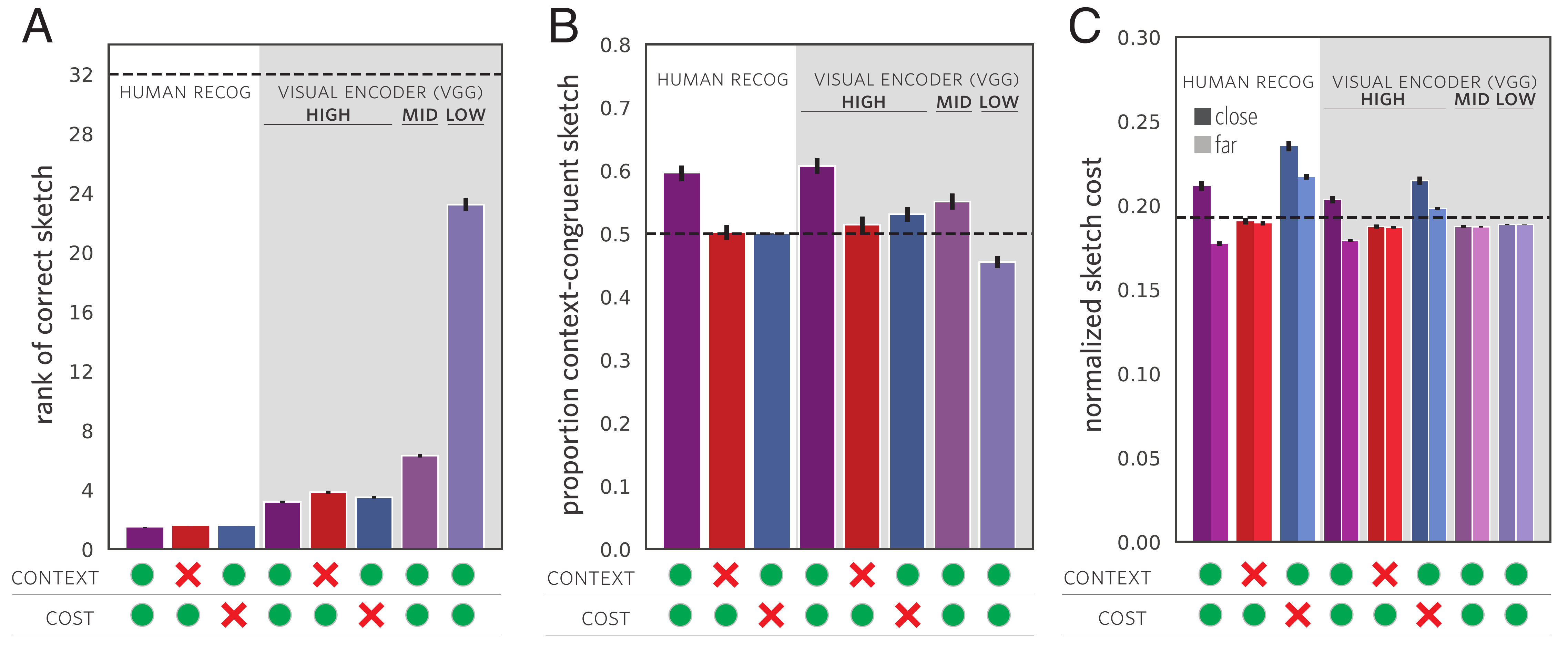}
\caption{Sketch production behavior by model variant. 
The table below each bar plot indicates whether the corresponding model variant plotted above is sensitive to context or cost. 
A green disc indicates context/cost sensitivity; a red `X' indicates the lack of context/cost sensitivity. 
Results in lefthand region of each panel (white background) reflect model predictions when using empirical estimates of $sim(s,o)$ based on human sketch recognition behavior. 
Results in the righthand region (gray background) reflect model predictions when using variants of the visual encoder that represented sketches and objects at varying levels of visual abstraction (i.e., high, mid, low).
All results reflect average model behavior on test data across five crossvalidation folds. 
Error bars represent 1 s.e. for this average estimate, found by applying inverse-variance weighting on individual confidence intervals from each train-test split. 
A: Rank of target sketch in list of 64 object-context categories, ordered by the probability assigned by each model. 
Dashed line reflects expected target rank under uniform guessing. 
Distribution of target rank scores across models suggest that high-quality estimates of $sim(s,o)$ are critical for strong performance. 
B: Proportion of trials on which each model assigned a higher rank to the context-congruent sketch of the correct object than the context-incongruent version of the correct object. 
Dashed line reflects expected behavior under indifference between the two versions of the sketch. 
Only models above this line show consistent and appropriate modulation of sketch producton by context. 
C: Normalized time cost of sketches produced by each model.
Predicted sketch cost on each trial computed by marginalizing over probabilities assigned to each sketch category. 
Darker bars reflect behavior in the close condition; lighter bars the far condition. 
Dashed line indicates the average cost of sketches in the full dataset; bars below this line reflect a preference for sketches that are less costly than average, bars above this line for sketches that are costlier than average. 
Only models that span this dashed line match the pattern of contextual modulation of sketch cost displayed by human sketchers.}
\label{model_results}
\end{figure*}

We found that the full model, $S_{prag}$, provided a much better fit to human behavior than the context-insensitive variant, $S_{sim}$ (median log Bayes Factor = $16.1$; see Table \ref{model_comparison}), and the cost-insensitive variant, $S_{nocost}$ (BF = $9.54$).
To gain further insight into the functional consequences of each lesion, we investigated three aspects of each model's behavior: (a) \textit{sketch retrieval}: the ability to assign a high absolute rank to the target sketch category in context, out of the 64 object-context alternatives; (b) \textit{context congruity}: the ability to consistently assign a higher rank to the context-congruent version of the target object over the context-incongruent version; and (c) \textit{cost modulation}: how consistently a model produced costlier sketches than average in the close condition, and less costly sketches than average in the far condition, mirroring human behavior.

We found that in general, sketch retrieval performance was high for all three model variants (target rank 95\% CI: pragmatic = $[1.43, 1.50]$, context-insensitive = $[1.54, 1.60]$, cost-insensitive = $[1.55, 1.60]$) (Fig.~\ref{model_results}A, left).
However, only the pragmatic sketcher was able to reliably produce the sketch appropriate for the context condition more frequently than would be predicted by chance (95\% CI proportion: $[0.571, 0.620]$; Fig.~\ref{model_results}B, left); neither the context-insensitive nor the cost-insensitive variants displayed this context congruity (95\% CI: context-insensitive = $[0.478, 0.525]$, cost-insensitive = $[0.498, 0.501]$). 
We observed that the lack of context congruity in the lesioned variants was attributable to an overall bias towards close sketches, which are highly informative in absolute terms, and thus higher in communicative utility if the distractors or sketch cost is ignored. 

Moreover, only the pragmatic sketcher produced costlier sketches than average in the close condition (95\% CI normalized cost: $[0.205, 0.218]$ vs. grand mean cost = $0.196$; Fig.~\ref{model_results}C, left), and less costly sketches than average in the far condition (95\%CI: $[0.175, 0.180]$). 
The context-insensitive variant is inherently unable to modulate the cost of the sketches it produces by context condition, and thus was no more or less likely to select a costlier, more diagnostic sketch on a close trial (95\% CI: $[0.187, 0.194]$) than a far trial (95\% CI: $[0.187, 0.192]$), and preferred slightly less costly sketches overall. 
While the cost-insensitive variant did exhibit cost modulation by context, because it ignores their cost, it preferred costlier sketches overall in both close (95\%CI: $[0.229, 0.241]$) and far contexts (95\%CI: $[0.214, 0.220]$). 
Together, these results suggest that both context and cost sensitivity are critical for capturing key aspects of contextual flexibility in human visual communication. 

\subsubsection*{Evaluating contribution of visual abstraction}

Having established the importance of pragmatic inference, we next sought to formally evaluate the contribution of visual abstraction.
Such formal evaluation requires an explicit model of how abstract perceptual information is extracted from raw visual inputs across successive stages of visual processing. 
Employing a model that operates directly on image inputs is important for a computational theory of visual communication because it allows our full model to generate predictions for novel sketches and contexts.

Leveraging recent advances in computational vision \cite{FanCommon2018,yamins2014performance}, we instantiated the visual encoder as a deep convolutional neural network (DCNN).
This choice of model class is motivated by prior work showing that such networks, in addition to being a type of universal function approximator \cite{hornik1991approximation}, learn higher-layer feature representations that capture more abstract perceptual information in drawings \cite{FanCommon2018}, capture perceptual judgments of object shape similarity \cite{kubilius2016deep}, and predict neural population responses in categories across the ventral visual stream \cite{yamins2014performance} when trained on challenging natural object recognition tasks \cite{deng2009imagenet}. 


Concretely, the visual encoder is a function that accepts a pair of images as input (see Fig. \ref{model_schematic}A): a sketch, $s$, and an object rendering, $o$, and returns a scalar value reflecting the degree of perceptual correspondence between the sketch and object, $\textrm{sim}(s,o)$, which lies in the range $[0,1]$, where $\textrm{sim}(s,o)=0$ reflects minimal correspondence and $\textrm{sim}(s,o)=1$ reflects maximal correspondence.
This encoder consists of two functional components: a base visual encoder network, $B$, and an adaptor network, $A$: $\textrm{sim}(s,o) = A(B(s,o))$.

We employ a widely used and high-performing deep convolutional neural network architeceture, VGG-19, pretrained to recognize objects from the Imagenet database as our base visual encoder, whose parameters remain frozen \cite{simonyan2014very}. 
We then augment the pretrained feature representation of the base encoder with a shallow adaptor network, which is trained to predict the perceptual correspondence between specific sketch-object pairs.
The reason we train an adaptor network is that although prior work has shown that representation of object \textit{categories} converges for sketches and photos at higher layers in DCNN models trained only on photos \cite{FanCommon2018}, additional supervision can substantially improve the accuracy of predictions involving comparisons between sketches and photos at the \textit{instance} level \cite{sangkloy2016sketchy}. 


To evaluate the importance of the greater visual abstraction available at higher layers of VGG-19, we compare adaptor networks that intercept VGG-19 image representations at three different layers: the first max pooling layer (early), the tenth convolutional layer (mid), and the first fully connected layer (high).
Each adaptor network was trained to predict the empirical estimates of sketch-object correspondence from the recognition experiment, and evaluated on held out data in a 5-fold crossvalidated manner using the same splits as in the previous section.

Consistent with our hypothesis, we found that a pragmatic sketcher model employing high-level features provided a substantially better fit to the data than one using mid-level features (high vs. mid median BF: $94.8$) or low-level features (high vs. low BF: $257$).
These results show that making fuller use of the depth of VGG to compute the perceptual correspondence between a sketch and object yields a stronger basis for explaining human visual communication behavior.
Unsurprisingly, this pragmatic sketcher model employing high-level features did not fit the data as well as the pragmatic sketcher model that could directly access human recognition data (median log Bayes Factor = $105.71$).
However, a major advantage of incorporating a visual encoder is the capacity to generalize to novel sketches without requiring the collection of additional recognition data for each new sketch.

To further probe the functional consequences of decreasing the capacity for visual abstraction, we investigated each model variant on sketch retrieval, context congruity, and cost modulation. 
Critically, we found that high-level features supported strong performance on sketch retrieval (95\% CI target rank: $[3.03, 3.37]$, Fig.~\ref{model_results}A), compared to mid-level features (target rank: $[6.05, 6.56]$) and low-level features (target rank: $[22.4, 24.1]$). 
These results show that without a high-performing visual encoder, the model is much less likely to produce sketches of the correct object, a basic prerequisite for successful visual communication even in the absence of contextual variability. 

Moreover, the pragmatic sketcher model using high-level features also displayed context congruity (95\% CI: $[0.583, 0.632]$, Fig.~\ref{model_results}B), comparable in degree to the best-performing pragmatic model that operated directly on empirical estimates of sketch-object correspondence, showing that our full sketcher model displayed this signature of contextual flexibility for novel communicative contexts and sketches. 
The variant using mid-level features also displayed context congruity to a weaker extent (95\% CI: $[0.526, 0.576]$), suggesting that an intermediate level of visual abstraction is sufficient to achieve an intermediate degree of context congruity. 
By contrast, the variant using low-level features failed to prefer the context-congruent sketch category (95\% CI: $[0.435, 0.475]$), providing a lower bound on the level of visual abstraction required in the underlying encoder to support flexible visual communication behavior. 

\newgeometry{margin=1.4in} 
\begin{landscape}

\begin{table}[]
\begin{tabular}{@{}lll|llll@{}}
\toprule
\multicolumn{1}{{c}}{} & \multicolumn{2}{c|}{\textbf{human recog}} & \multicolumn{4}{c}{\textbf{visual encoder}} \\ \midrule
\multicolumn{1}{c}{\textbf{split}} & \multicolumn{1}{c}{\textbf{context vs. no-context}} & \multicolumn{1}{c|}{\textbf{cost vs. no-cost}} & \multicolumn{1}{c}{\textbf{context vs. no-context}} & \multicolumn{1}{c}{\textbf{cost vs. no-cost}} & \multicolumn{1}{c}{\textbf{high vs. mid}} & \multicolumn{1}{c}{\textbf{high vs. low}} \\
\textbf{1} & 18.0 & 11.9 & 44.5 & 2.70 & 105 & 282 \\
\textbf{2} & 8.46 & 9.89 & 20.9 & -0.33 & 92.5 & 242 \\
\textbf{3} & 19.2 & 8.95 & 31.9 & 1.98 & 94.8 & 257 \\
\textbf{4} & 13.4 & 9.54 & 8.35 & -0.67 & 93.4 & 248 \\
\textbf{5} & 16.1 & 7.92 & 28.1 & 5.99 & 114 & 269 \\
\textbf{median} & \textbf{16.1} & \textbf{9.54} & \textbf{28.1} & \textbf{1.98} & \textbf{94.8} & \textbf{257} \\ \bottomrule
\end{tabular}
\caption{Log Bayes Factors (BF) for comparisons between full and lesioned model variants (columns) for each crossvalidation fold (rows). 
Log-BFs$>$0 indicate greater evidence for the full model than the lesioned variant. 
Columns under the human recog header contain comparisons between model variants that used empirical estimates of perceptual correspondence based on human sketch recognition behavior. 
Columns under the visual encoder header contain comparisons between model variants that used a deep convolutional neural network visual encoder, trained in a five-fold crossvalidated manner using human sketch recognition behavior. 
The context vs. no-context columns includes comparisons between context-sensitive and context-insensitive variant;
the cost vs. no-cost columns includes comparisons between cost-sensitive and cost-insensitive variant; 
the high vs. mid column includes comparisons between model variants using a high adaptor vs. mid adaptor in a context/cost-sensitive model; 
and the high vs. low column inclues comparisons between between model variants using a high adaptor vs. low adaptor in context/cost-sensitive model.}
\label{model_comparison}
\end{table}

\end{landscape}
\restoregeometry

Again, only the pragmatic sketcher model using high-level features displayed the same qualitative pattern of cost modulation as people did (95\%CI: close = $[0.199, 0.208]$, far = $[0.178, 0.181]$, Fig.~\ref{model_results}C), while both of the other variants using mid-level and low-level features failed to do so (95\%CI: mid-level: close = $[0.186, 0.189]$, far = $[0.186, 0.188]$; low-level: close = $[0.188, 0.189]$, far = $[0.188, 0.189]$).

These results so far show the best-performing visual encoder to be the one making fuller use of the depth of the base visual encoder to extract more abstract perceptual properties, providing strong evidence for the importance of a high degree of visual abstraction for explaining our empirical findings. 
Next, we performed the same context and cost sensitivity lesion experiments as before in order to evaluate the contribution of pragmatic inference in our full sketcher model. 
Again, we found that the pragmatic sketcher provided a stronger overall fit to human behavior than the context-insensitive variant (median log Bayes Factor = $28.1$; see Table \ref{model_comparison}), and a modestly better fit than the cost-insensitive variant (BF = $1.98$). 
Critically, we found that removing context and cost sensitivity diminished the ability of this model to produce the context-congruent sketch of the correct object (context-insensitive 95\% CI: $[0.489, 0.539]$; cost-insensitive 95\% CI: $[0.507, 0.554]$; Fig.~\ref{model_results}B), and appropriately modulate the cost of the sketches it produced (context-insensitive 95\% CI: close = $[0.185, 0.190]$, far = $[0.185, 0.189]$; cost-insensitive 95\% CI: close = $[0.210, 0.219]$, far = $[0.196, 0.200]$; Fig.~\ref{model_results}C). 
By contrast, these lesions led to only modest decrements in overall sketch retrieval performance (95\% CI target rank: context-insensitive = $[3.65, 4.05]$, cost-insensitive = $[3.33, 3.67]$;  Fig.~\ref{model_results}A), suggesting that the visual encoder itself is a major determinant of the ability to produce sketches of the correct \textit{object}, even if not the context-congruent version.
These results converge with those of the lesion experiments conducted on the pragmatic sketcher model lacking a visual encoder, and together provide strong evidence for the importance of both visual abstraction and pragmatic inference for explaining contextual flexibility in human visual communication. 



\section*{Discussion}

The present study examined how communicative context influences visual communication behavior in a drawing-based reference game. 
We explored the hypothesis that people spontaneously account for information in common ground with their communication partner to produce drawings that are diagnostic of the target relative to the alternatives, while not being too costly to produce. 
We found that people spontaneously modulate how much time they invest in their drawings according to how similar the distractors are to the target, spending more time to produce more informative sketches when the alternatives were highly similar, but getting away with spending less time and producing less informative drawings when the alternatives were highly distinct.
Observing such contextual flexibility provides strong evidence that visual communication about objects is not constrained exclusively by the visual properties of that object alone.  
Rather, our findings expose a critical role for pragmatic inference --- the ability to infer what information would not only be true, but be \textit{relevant} to communicate in context.
To test this hypothesis, we developed a computational model that embodied both pragmatic inference and visual abstraction, and found that it predicted human communication behavior well, and outperformed variants of the model lacking either component. 
Together, this paper provides a first algorithmically explicit theory of how visual perception and social cognition support contextual flexibility during visual communication.

This work generalizes the Rational Speech Act (RSA) modeling framework, originally developed to explain contextual effects in verbal communication \cite{frank2012predicting,goodman2013knowledge,franke2016probabilistic,bergen2016pragmatic}, to the domain of visual communication.
RSA models take inspiration from the insights of Paul Grice \cite{grice1975syntax}, and incorporate ideas from decision theory, probabilistic models of cognition, bounded rationality, and linguistics, to understand how substantial variance in natural language use can be explained by general principles of social cognition. 
They have been shown to capture key patterns of natural language use \cite{goodman2013knowledge}, achieve good quantitative fits with experimental data \cite{kao2014formalizing}, and enhance the ability of artificial agents to produce informative language in reference game tasks \cite{monroe2017colors,Cohn-GordonGP18}.
In extending this modeling framework to the visual domain, our findings provide novel evidence for the possibility that similar cognitive mechanisms may underlie pragmatic behavior across different communication modalities, a notion implicitly endorsed by prior work that has used non-linguistic modalities to investigate general constraints on communication 
\cite{goldin1977development,Garrod:2007wk,fay2010interactive,theisen2010systematicity,garrod2010can,Galantucci:2005uh,verhoef2014emergence}. 
Moreover, these findings provide a functional account of how drawings spanning widely varying amounts of detail can all nevertheless be effective carriers of meaning, depending on how much and what kind of information is shared between communicators. 
A fruitful avenue for future research would be to augment the current model with the capacity to accumulate such shared information over time, potentially endowing it with the capacity to develop conventionalized ways of depicting objects that are increasingly efficient, at least within a specific context \cite{Garrod:2007wk}. 




There are several limitations of our model that would be valuable to address in future work. 
First, obtaining a visual encoder that could produce accurate predictions of perceptual correspondence between sketch-object pairs required substantial supervision. 
While heavy supervision is not uncommon when developing neural network models of sketch representation \cite{sangkloy2016sketchy,yu2017sketch,song2017deep}, future work should investigate architectures that require weaker supervision to estimate image-level correspondences between sketches and natural photographs. 
One promising approach may be to exploit the hierarchical and compositional structure of natural objects (i.e., parts, subparts, and their relations), as they are expressed in both natural images and sketches of objects \cite{battaglia2016interaction,mrowca2018graph}.
Second, our model produces a decision over which \textit{type} of sketch to produce in context, rather than producing a \textit{particular} sketch.  
This is of course different from the action selection problem human participants face --- they must decide not only what stroke to make, but where to place them, how many, and in what order.
While there have been recent and promising advances in modeling sketch production as a sequence of such actions \cite{lake2015human,ha2017neural,ganin2018synthesizing}, these approaches have not yet been shown to successfully emulate how people sketch real objects, much less how this behavior is modulated by communicative context. 
Future work should develop sketch production models that both operate on natural visual inputs and more closely approximate the the action space inherent to the task.
Meeting these challenges is not only important for developing more human-like artificial intelligence, but may also shed new light on the nature of human visual abstraction, and how ongoing perception and long-term conceptual knowledge guide action selection during complex, natural behaviors. 

In the long term, investigating the computational basis of visual communication may shed light on the sources of cultural variation in pictorial style, and lead to enhanced interactive visualization tools for education and research.

\section*{Materials and Methods}

\subsection*{Communication experiment: Manipulation of context in sketch-based reference game}

\subsubsection*{Participants}
A total of 192 unique participants, who were recruited via Amazon Mechanical Turk (AMT) and grouped into pairs, completed the experiment. 
They were provided a base compensation of \$2.70 for participation and earned a \$0.03 bonus for each correct trial. 
In this and subsequent behavioral experiments, participants provided informed consent in accordance with the Stanford IRB.
\subsubsection*{Stimuli and Task}
Because our goal was to understand how context influences the level of detail people use to distinguish objects from one another during visual communication, we populated our reference game with contexts possessing two key properties: (1) they contained familiar real-world objects, so that a primary source of variation would be driven by context, rather than difficulty recognizing or sketching the objects, \textit{per se}; and (2) they systematically varied in target-distractor similarity within a session, lending greater statistical power to comparisons between context conditions. 
To satisfy these objectives, we obtained 32 3D mesh models of objects belonging to 4 basic-level categories (i.e., birds, chairs, cars, dogs), containing eight objects each. 
Each object was rendered in color on a gray background at three-quarter perspective, 10$^{\circ}$ viewing angle (i.e., slightly above), and fixed distance. 
Independently in each experimental session, objects were allocated to eight sets of four objects: Four of these sets contained objects from the same category (``close''); the other four of these sets contained objects from different categories (``far'' condition).
The assignment of objects to set and condition was randomized across pairs.
Each set of four objects was presented four times each, such that each object in the quartet served as the target exactly once. 

Sketchers drew using black ink on digital canvas (pen width = 5 pixels; 300 x 300 pixels) embedded in a web browser window using Paper.js (http://paperjs.org/). Participants drew using the mouse cursor, and were not able to delete previous strokes. Each stroke of which was rendered on the viewer's screen immediately upon the completion of each stroke. There were no restrictions on how long participants could take to make their drawings. After clicking a submit button, the viewer guessed the identity of the drawn object by clicking one of the four objects in the array. Otherwise, the viewer had no other means of communicating with the sketcher. Both participants received immediate task-related feedback: the sketcher learned which object the viewer had clicked, and the viewer learned the identity of the target. Both participants earned bonus points for each correct response.


\subsection*{Recognition experiment: Measuring perceptual similarity between sketches and objects}

\subsubsection*{Participants}

A total of 112 participants were recruited via Amazon Mechanical Turk (AMT). They were provided a base compensation of \$1.00 for their participation, and earned an additional \$0.01 bonus for each correct response.

\subsubsection*{Task}
On each trial, participants were presented with a randomly selected sketch collected in the communication experiment, surrounded by a grid containing the 32 objects from that experiment. 
Their goal was to select the object in the grid that best matched the sketch. 
Participants received task feedback in the form of a bonus earned for each correct trial. 
Participants were instructed to prioritize accuracy over speed. 
We applied a conservative outlier removal procedure based on response latency, whereby trials that were either too fast to have supported careful consideration of the sketch and menu of objects (RT<$1000ms$), or too slow and suggestive of an attentional lapse (RT>$30s$), were filtered from the dataset. 
The removal of these outlier trials ($8.01$\%) did not have a substantial impact on the pattern of recognition behavior. 
In order to mitigate the possibility that participants could adjust their matching strategy according to any particular sketcher's style, each session was populated with 64 sketches sampled randomly from different reference games. 
To obtain robust estimates of sketch-object perceptual correspondences, each sketch was presented approximately 10 times across different sessions.  


\subsection*{Computational modeling}



\subsubsection*{Sketch data preprocessing} 
To train and evaluate our sketcher model, we first filter the sketch dataset to retain only sketches that were correctly identified by the viewer during the communication task (6.2\% incorrect) and were compliant with task instructions by not including `drawn' text annotations (4.4\% non-compliant). 
This filtered sketch dataset was then split into training, validation, and test sets in a 80\%, 10\%, and 10\% ratio, and this split was performed in a 5-fold crossvalidated manner.
Splits were based on context, defined as the set containing a specific target object and three distractor objects, such that no context appeared both in the training and test splits of any cross-validation fold. 
We ensured that: (1) the number of sketches from each category (i.e. car) and (2) the proportion of sketches from close and far trials were equated across splits. 
This was done to control for biases in model peformance due to imbalances in the training or test set.

\subsubsection*{Deriving empirical estimates of perceptual correspondence between sketches and objects}

In the recognition experiment, most sketches were not matched exclusively to a single object, but to several. 
We treated these sketches as thus displaying some degree of correspondence to the several objects it was matched to at least once. 
For a single sketch, we estimate the perceptual correspondence between that sketch and any object as the proportion of recognition task trials on which it was matched to that object. 
For sketches in each of 64 object-context categories, we estimate the ``aggregated'' sketch-object correspondence to be the proportion of recognition task trials on which any sketch from this category was matched to that object. 
Because our goal was to understand how well each model could produce produce informative sketches according to the context condition, and not necessarily to reproduce exactly the same sketch as a particular participant had on a specific trial, we use this aggregate correspondence measure in all modeling experiments.  
As a result, sketch-object correspondence scores lie in the range $[0,1]$, and sum to 1 for sketches in the same object-context category. 
Because all sketches from the same object-context category share the same correspondence to each object, there are a total of 32 sketch categories x 32 objects x 2 contexts = 2048 empirical perceptual correspondence scores.


\subsubsection*{Deriving empirical estimates of sketch costs}

We reasoned that the amount of time taken producing each sketch would be a natural proxy for the cost incurred by workers on Amazon Mechanical Turk, who increase their total compensation by completing tasks in a timely manner. 
However, as there were no absolute constraints on the amount of time that could be spent on each trial, there was considerable variability across different participants in terms of how much time they spent producing their sketches. 
To control for this variability across participants and to ensure robust estimates, we first removed outliers (draw times exceeding 5 s.d. from the mean), then z-score normalized drawing times across all remaining trials within a participant, and finally averaged these normalized draw times across sketches within the same object-context category as above, yielding 32 objects x 2 contexts = 64 empirical cost estimates in total.

\subsubsection*{Visual encoder architecture}

The visual encoder is a function that accepts a pair of images (both 224 x 224 RGB)---a sketch and an object rendering---as input and returns a scalar value in the range $[0,1]$, reflecting the degree of perceptual correspondence between the sketch and object. 

The encoder consists of two components: a base visual encoder and an adaptor network. 
We employed VGG-19 \cite{simonyan2014very} as our base visual encoder architecture.
We augmented VGG-19 with a shallow fully-connected \textit{adaptor} network that is trained to predict the perceptual correspondence between individual sketch-objecet pairs. 
Only the parameters of this adaptor network are trained; the parameters of the base visual encoder remain frozen. 
We compared three adaptor networks that intercept VGG-19 image representations at different layers: the first max pooling layer (early), the tenth convolutional layer (mid), and the first fully connected layer (high). 
To facilitate comparison between adaptor networks, we ensured that each of the three adaptors contain a comparable number of trainable parameters (number of learnable parameters for high: $1048839$; mid: $1049115$; low: $1048833$) with identical training hyperparameters (i.e., learning rate, batch size, etc.). 
To discriminate which layer provides the best starting feature basis for predicting sketch-object correspondence, these adaptor networks were also deliberately constrained to be shallow, i.e., consisting only of two linear layers with an intervening point-wise nonlinearity.

\textbf{High.} When applying the high-level visual encoder, a sketch and object were first passed through VGG and a feature vector in $\mathbb{R}^{4096}$ for each image is extracted from the one of the highest layers (i.e., the first fully-connected layer, also known as \textit{fc6}). 
These two vectors were then concatenated to form a single vector in $\mathbb{R}^{8192}$, to be passed into the high adaptor network. 
The high adaptor is composed of one linear layer that maps from $\mathbb{R}^{8192} \rightarrow \mathbb{R}^{128}$, followed by a ``Swish'' nonlinearity \cite{ramachandran2018searching} and dropout, then a second linear layer mapping from $\mathbb{R}^{128} \rightarrow \mathbb{R}^{1}$.
Swish is a recently discovered nonlinearity that outperforms the common rectified linear nonlinearity (ReLU) in deep models on several benchmarks \cite{ramachandran2018searching}.
Dropout was applied to mitigate overfitting and improve generalization \cite{hinton2012improving,gal2015dropout}.

\textbf{Mid.} When applying the mid-level visual encoder, sketch and object representations are intercepted from an intermediate layer (i.e., the 10th convolutional layer, \textit{conv\_4\_2}).
Features in this layer are of dimensionality 512 x 28 x 28.
Each of the sketch and object feature tensors were then ``flattened'' to a one dimensional vector in $\mathbb{R}^{512}$ using a weighted linear combination over the spatial dimensions $\sum_{i=1}^{28}\sum_{j=1}^{28} w_{ij} * x_{ij}$, where $x_{ij}$ indexes a spatial location in the image representation at this layer (i.e., `soft attention' over the spatial dimension, \cite{xu2015show}). 
These weights $\{w_{ij}|1\leq i,j \leq 28\}$ are learned jointly with the parameters of the rest of the mid adaptor, but learned independently between sketch and object image modalities \cite{xu2015show}. 

The two feature vectors in $\mathbb{R}^{512}$ are then concatenated to form a single vector in $\mathbb{R}^{1024}$.
Following the architecture of the high adaptor, the mid adaptor consists of a linear layer that maps from $\mathbb{R}^{1024} \rightarrow \mathbb{R}^{1021}$, followed by a Swish nonlinearity, dropout, then a linear layer from $\mathbb{R}^{1021} \rightarrow \mathbb{R}^{1}$. 

\textbf{Low.} When applying the low-level visual encoder, sketch and object representations are intercepted from the first max pooling layer (i.e., \textit{pool1}).
Features in this layer are of dimensionality 64 by 112 by 112. 
As above, a weighted sum of model activations over the spatial dimension was applied first (112 x 112), yielding a sketch and object vector, both in $\mathbb{R}^{64}$, which were then concatenated to form a single vector in $\mathbb{R}^{128}$. 
This was followed by a linear layer that maps from $\mathbb{R}^{128} \rightarrow \mathbb{R}^{7875}$, then a Swish nonlinearity, dropout, and a final linear layer that maps from $\mathbb{R}^{7875} \rightarrow \mathbb{R}^{1}$. 

The penultimate hidden layer sizes in the mid (i.e., 1021 units) and low adaptors (i.e., 7875 units) were chosen to ensure that the total number of learnable parameters matched the high adaptor as closely as possible. 


\subsubsection*{Visual encoder training}

We trained each adaptor (i.e., high, mid, low) to predict, for each sketch, a 32-dimensional vector that captures the \textit{pattern} of perceptual correspondences between that sketch and all 32 objects. 
Each encoder accepts a sketch-object pair as input and returns a real number as output, 
reflecting their perceptual correspondence.
We iterate over all objects in the stimulus set $\mathcal{I}$ to generate the predicted 32-vector for each sketch, and then apply softmax normalization, yielding a vector that sums to 1. 
We define the loss function, $\mathcal{L}$, to be the cross entropy loss between the predicted distribution, $q$ and the empirically estimated perceptual correspondence vector, $p$ (which also sums to 1):

\begin{equation}
    \mathcal{L} = \sum_{x \in \mathcal{I}} p(x)\log q(x)
    \label{eqn:cross_entropy}
\end{equation}


This loss function explicitly encourages the adaptor to learn not only to predict the strength of the correspondence between a sketch and the object it was intended to depict (measured by correct matches during recognition), but also to predict its correspondence to all of the other objects (measured by the pattern of confusions during recognition).

We use the Adam optimization algorithm \cite[]{kingma2014adam} (learning rate = 1e-4) over minibatches of size 10 for 100 epochs, where an epoch is a full pass through the training set.\footnote{As a property of the input domain, the gradients with respect to adaptor parameters are very small (1.51e-4 $\pm$ 2.61e-4), inevitably resulting in poor learning (we can reproduce this effect from several intializations). We find that naively increasing the learning rate led to unstable optimization, but that multiplying the loss by a large constant $C$ leads to a much smoother learning trajectories and good test generalization. Critically, increasing the learning rate and multiplying the loss by a constant are not equivalent for second moment gradient methods. In practice, $C =$ 1e4.}  
After training each adaptor for 100 epochs, we select the model found during training with the best performance on the validation set. 

\subsubsection*{Generating encoder-based estimates of perceptual correspondence between sketches and objects}

To generate sketch-object correspondence scores for sketches in each test split, we first pass each sketch-object pair into a visual encoder, yielding a single image-level correspondence score lying in the range $(-\infty,+\infty)$. 
To map these raw image-level scores to the appropriate range for a correspondence score ($[0,1]$), we first z-score them ($f(x) = \frac{x - \bar{x}}{\mathrm{s}}$), then apply the logistic function ($f(x)= \frac{1}{1+e^{-x}}$).
These normalized image-level correspondence scores are then averaged across all sketches belonging to the same object-context category, yielding 32 objects x 32 sketches x 2 contexts  = 2048 model-based perceptual correspondence scores for each visual encoder variant (i.e., high, mid, low).

\subsubsection*{Model comparison}

In order to test the contribution of each component of our sketcher model, we conducted a series of lesion experiments and formal model comparisons.
To quantify the evidence for one model over another, we computed Bayes Factors:
the ratio of likelihoods for each model, integrating over all their respective parameters under the prior:
$$BF = \frac{\int P(D | M_1, \theta_1)P(\theta_1)}{\int P(D | M_2, \theta_2)P(\theta_2)}$$
Unlike classical likelihood ratio tests, which use the maximum likelihood, the Bayes Factor naturally penalizes models for their complexity \cite{wagenmakers2018bayesian,jefferys1992ockham}.
We placed uninformative uniform priors over all five parameters required to specify our models: a discrete choice over alternative approaches to computing perceptual correspondance: 
$$m \sim \textrm{Unif}\{\textrm{``human recog''}, \textrm{``high''}, \textrm{``mid''}, \textrm{``low''}\}$$
and over the continuous latent parameters, 
$$w_i, w_c, w_d, \alpha \sim \textrm{Unif}(0, 50).$$ 
To compute the likelihood function $P(D | M, \theta)$ for a speaker model $M$ under parameters $\theta$, we perform exact inference for our sketcher model using (nested) enumeration and sum over all test set datapoints within a crossvalidation fold. 



Specifically, we compute the exact likelihood at every point on a discrete grid of parameters.
This is of particular interest for nested model comparisons, e.g. comparing our full model to a context-insensitive variant.
Rather than computing the full marginalized likelihood for both models, we can use the Savage-Dickey method \cite{wagenmakers2010bayesian} to simply compare the posterior probability against the prior at the nested point of interest (e.g. $w_c = 0$) for the full model.

To evaluate the contribution of pragmatic inference, we begin by comparing the pragmatic sketcher model using empirically estimated perceptual correspondences to nested ``cost-insensitive'' ($w_c = 0$) and ``context-insensitive'' ($w_d = 0$) variants. To evaluate the contribution of visual abstraction, we then proceed to compare the three visual encoder variants that adapt features from different layers of VGG-19, marginalizing over all other parameters. Finally, we perform the same context and cost lesion experiments on the full model that employed the best-performing visual encoder (i.e., ``high'').

\subsubsection*{Evaluating model predictions}


We implemented our models and conducted inference in the probabilistic programming language WebPPL \cite{goodman2014design}.
We use MCMC to draw 1000 samples from the joint posterior with a lag of 0, discarding 3000 burn-in samples.
We constructed posterior predictive distributions by computing each measure of interest (i.e., target rank, context congruity, sketch cost) over the test data set, for every MCMC sample.
To estimate standard errors on predictions across models, we employed the following procedure to account for three nested sources of variation: variation across trials within a test split, variation across the parameter posterior within a test split, and variation across test splits.  
Specifically, for each model variant and for each test split we bootstrap resampled trials with replacement from the test dataset 1000 times to estimate the mean and standard error on each measure of interest, marginalizing over MCMC samples from the parameter posterior. 
We applied inverse-variance weighting to aggregate these estimates of the mean and standard error across test splits, such that test splits with lower variance contribute more than do splits with higher variance, yielding an overall estimate of the mean and standard error for each measure of interest, for each model variant. 
We estimated the half-widths of the 95\% confidence interval for each measure of interest under the the assumption of normality for the sampling distribution of the mean.

\subsubsection*{Code availability} The code for the analyses presented in this article is publicly available in a Github repository at: \url{https://github.com/judithfan/visual_communication_in_context}.

\subsubsection*{Data availability} The data presented in this article will be made publicly available in a figshare repository upon acceptance.
\section*{Acknowledgements}

Thanks to Dan Yamins and the Stanford CoCo Lab for helpful comments and discussion.

\section*{Author contributions statement}

J.E.F and R.X.D.H. designed and conducted human experiments, J.E.F, R.X.D.H, and M.W. analyzed data and performed computational modeling. J.E.F, R.X.D.H, M.W., and N.D.G. formulated models, interpreted results, and wrote the paper.

\section*{Additional information}

The authors declare no competing interests.

\bibliographystyle{apacite}

\setlength{\bibleftmargin}{.125in}
\setlength{\bibindent}{-\bibleftmargin}

\newpage
\bibliography{references}

\begin{thebibliography}{}

\bibitem [\protect \citeauthoryear {%
Abell%
}{%
Abell%
}{%
{\protect \APACyear {2009}}%
}]{%
abell2009canny}
\APACinsertmetastar {%
abell2009canny}%
\begin{APACrefauthors}%
Abell, C.%
\end{APACrefauthors}%
\unskip\
\newblock
\APACrefYearMonthDay{2009}{}{}.
\newblock
{\BBOQ}\APACrefatitle {Canny resemblance} {Canny resemblance}.{\BBCQ}
\newblock
\APACjournalVolNumPages{Philosophical Review}{118}{2}{183--223}.
\PrintBackRefs{\CurrentBib}

\bibitem [\protect \citeauthoryear {%
Aubert%
\ \protect \BOthers {.}}{%
Aubert%
\ \protect \BOthers {.}}{%
{\protect \APACyear {2014}}%
}]{%
Aubert:2014jy}
\APACinsertmetastar {%
Aubert:2014jy}%
\begin{APACrefauthors}%
Aubert, M.%
, Brumm, A.%
, Ramli, M.%
, Sutikna, T.%
, Saptomo, E\BPBI W.%
, Hakim, B.%
\BDBL {}Dosseto, A.%
\end{APACrefauthors}%
\unskip\
\newblock
\APACrefYearMonthDay{2014}{{\APACmonth{09}}}{}.
\newblock
{\BBOQ}\APACrefatitle {{Pleistocene cave art from Sulawesi, Indonesia}}
  {{Pleistocene cave art from Sulawesi, Indonesia}}.{\BBCQ}
\newblock
\APACjournalVolNumPages{Nature}{514}{7521}{223--227}.
\PrintBackRefs{\CurrentBib}

\bibitem [\protect \citeauthoryear {%
Battaglia%
, Pascanu%
, Lai%
, Rezende%
\BCBL {}\ \protect \BOthers {.}}{%
Battaglia%
\ \protect \BOthers {.}}{%
{\protect \APACyear {2016}}%
}]{%
battaglia2016interaction}
\APACinsertmetastar {%
battaglia2016interaction}%
\begin{APACrefauthors}%
Battaglia, P.%
, Pascanu, R.%
, Lai, M.%
, Rezende, D\BPBI J.%
\BCBL {}\ \BOthersPeriod {.}\end{APACrefauthors}%
\unskip\
\newblock
\APACrefYearMonthDay{2016}{}{}.
\newblock
{\BBOQ}\APACrefatitle {Interaction networks for learning about objects,
  relations and physics} {Interaction networks for learning about objects,
  relations and physics}.{\BBCQ}
\newblock
\BIn{} \APACrefbtitle {{Advances in Neural Information Processing Systems}}
  {{Advances in Neural Information Processing Systems}}\ (\BPGS\ 4502--4510).
\PrintBackRefs{\CurrentBib}

\bibitem [\protect \citeauthoryear {%
Bergen%
, Levy%
\BCBL {}\ \BBA {} Goodman%
}{%
Bergen%
\ \protect \BOthers {.}}{%
{\protect \APACyear {2016}}%
}]{%
bergen2016pragmatic}
\APACinsertmetastar {%
bergen2016pragmatic}%
\begin{APACrefauthors}%
Bergen, L.%
, Levy, R.%
\BCBL {}\ \BBA {} Goodman, N.%
\end{APACrefauthors}%
\unskip\
\newblock
\APACrefYearMonthDay{2016}{}{}.
\newblock
{\BBOQ}\APACrefatitle {Pragmatic reasoning through semantic inference}
  {Pragmatic reasoning through semantic inference}.{\BBCQ}
\newblock
\APACjournalVolNumPages{Semantics and Pragmatics}{9}{}{}.
\PrintBackRefs{\CurrentBib}

\bibitem [\protect \citeauthoryear {%
Boltz%
}{%
Boltz%
}{%
{\protect \APACyear {1994}}%
}]{%
boltz1994origin}
\APACinsertmetastar {%
boltz1994origin}%
\begin{APACrefauthors}%
Boltz, W\BPBI G.%
\end{APACrefauthors}%
\unskip\
\newblock
\APACrefYear{1994}.
\newblock
\APACrefbtitle {The origin and early development of the Chinese writing system}
  {The origin and early development of the chinese writing system}\ (\BVOL~78).
\newblock
\APACaddressPublisher{}{Eisenbrauns}.
\PrintBackRefs{\CurrentBib}

\bibitem [\protect \citeauthoryear {%
Cohn{-}Gordon%
, Goodman%
\BCBL {}\ \BBA {} Potts%
}{%
Cohn{-}Gordon%
\ \protect \BOthers {.}}{%
{\protect \APACyear {2018}}%
}]{%
Cohn-GordonGP18}
\APACinsertmetastar {%
Cohn-GordonGP18}%
\begin{APACrefauthors}%
Cohn{-}Gordon, R.%
, Goodman, N\BPBI D.%
\BCBL {}\ \BBA {} Potts, C.%
\end{APACrefauthors}%
\unskip\
\newblock
\APACrefYearMonthDay{2018}{}{}.
\newblock
{\BBOQ}\APACrefatitle {Pragmatically Informative Image Captioning with
  Character-Level Inference} {Pragmatically informative image captioning with
  character-level inference}.{\BBCQ}
\newblock
\BIn{} \APACrefbtitle {Proceedings of the 2018 Conference of the North American
  Chapter of the Association for Computational Linguistics: Human Language
  Technologies, NAACL-HLT, New Orleans, Louisiana, USA, June 1-6, 2018, Volume
  2 (Short Papers)} {Proceedings of the 2018 conference of the north american
  chapter of the association for computational linguistics: Human language
  technologies, naacl-hlt, new orleans, louisiana, usa, june 1-6, 2018, volume
  2 (short papers)}\ (\BPGS\ 439--443).
\newblock
\begin{APACrefURL} \url{https://aclanthology.info/papers/N18-2070/n18-2070}
  \end{APACrefURL}
\PrintBackRefs{\CurrentBib}

\bibitem [\protect \citeauthoryear {%
Deng%
\ \protect \BOthers {.}}{%
Deng%
\ \protect \BOthers {.}}{%
{\protect \APACyear {2009}}%
}]{%
deng2009imagenet}
\APACinsertmetastar {%
deng2009imagenet}%
\begin{APACrefauthors}%
Deng, J.%
, Dong, W.%
, Socher, R.%
, Li, L\BHBI J.%
, Li, K.%
\BCBL {}\ \BBA {} Fei-Fei, L.%
\end{APACrefauthors}%
\unskip\
\newblock
\APACrefYearMonthDay{2009}{}{}.
\newblock
{\BBOQ}\APACrefatitle {Imagenet: A large-scale hierarchical image database}
  {Imagenet: A large-scale hierarchical image database}.{\BBCQ}
\newblock
\BIn{} \APACrefbtitle {{Computer Vision and Pattern Recognition, 2009.}}
  {{Computer Vision and Pattern Recognition, 2009.}}\ (\BPGS\ 248--255).
\PrintBackRefs{\CurrentBib}

\bibitem [\protect \citeauthoryear {%
Donald%
}{%
Donald%
}{%
{\protect \APACyear {1991}}%
}]{%
donald1991origins}
\APACinsertmetastar {%
donald1991origins}%
\begin{APACrefauthors}%
Donald, M.%
\end{APACrefauthors}%
\unskip\
\newblock
\APACrefYear{1991}.
\newblock
\APACrefbtitle {Origins of the modern mind: Three stages in the evolution of
  culture and cognition} {Origins of the modern mind: Three stages in the
  evolution of culture and cognition}.
\newblock
\APACaddressPublisher{}{Harvard University Press}.
\PrintBackRefs{\CurrentBib}

\bibitem [\protect \citeauthoryear {%
Fan%
, Yamins%
\BCBL {}\ \BBA {} Turk-Browne%
}{%
Fan%
\ \protect \BOthers {.}}{%
{\protect \APACyear {2018}}%
}]{%
FanCommon2018}
\APACinsertmetastar {%
FanCommon2018}%
\begin{APACrefauthors}%
Fan, J\BPBI E.%
, Yamins, D\BPBI L\BPBI K.%
\BCBL {}\ \BBA {} Turk-Browne, N\BPBI B.%
\end{APACrefauthors}%
\unskip\
\newblock
\APACrefYearMonthDay{2018}{}{}.
\newblock
{\BBOQ}\APACrefatitle {Common Object Representations for Visual Production and
  Recognition} {Common object representations for visual production and
  recognition}.{\BBCQ}
\newblock
\APACjournalVolNumPages{Cognitive Science}{0}{0}{}.
\newblock
\begin{APACrefURL}
  \url{https://onlinelibrary.wiley.com/doi/abs/10.1111/cogs.12676}
  \end{APACrefURL}
\newblock
\begin{APACrefDOI} \doi{10.1111/cogs.12676} \end{APACrefDOI}
\PrintBackRefs{\CurrentBib}

\bibitem [\protect \citeauthoryear {%
Fay%
, Garrod%
, Roberts%
\BCBL {}\ \BBA {} Swoboda%
}{%
Fay%
\ \protect \BOthers {.}}{%
{\protect \APACyear {2010}}%
}]{%
fay2010interactive}
\APACinsertmetastar {%
fay2010interactive}%
\begin{APACrefauthors}%
Fay, N.%
, Garrod, S.%
, Roberts, L.%
\BCBL {}\ \BBA {} Swoboda, N.%
\end{APACrefauthors}%
\unskip\
\newblock
\APACrefYearMonthDay{2010}{}{}.
\newblock
{\BBOQ}\APACrefatitle {The interactive evolution of human communication
  systems} {The interactive evolution of human communication systems}.{\BBCQ}
\newblock
\APACjournalVolNumPages{Cognitive Science}{34}{3}{351--386}.
\PrintBackRefs{\CurrentBib}

\bibitem [\protect \citeauthoryear {%
Frank%
\ \BBA {} Goodman%
}{%
Frank%
\ \BBA {} Goodman%
}{%
{\protect \APACyear {2012}}%
}]{%
frank2012predicting}
\APACinsertmetastar {%
frank2012predicting}%
\begin{APACrefauthors}%
Frank, M\BPBI C.%
\BCBT {}\ \BBA {} Goodman, N\BPBI D.%
\end{APACrefauthors}%
\unskip\
\newblock
\APACrefYearMonthDay{2012}{}{}.
\newblock
{\BBOQ}\APACrefatitle {Predicting pragmatic reasoning in language games}
  {Predicting pragmatic reasoning in language games}.{\BBCQ}
\newblock
\APACjournalVolNumPages{Science}{336}{6084}{998--998}.
\PrintBackRefs{\CurrentBib}

\bibitem [\protect \citeauthoryear {%
Franke%
\ \BBA {} J{\"a}ger%
}{%
Franke%
\ \BBA {} J{\"a}ger%
}{%
{\protect \APACyear {2016}}%
}]{%
franke2016probabilistic}
\APACinsertmetastar {%
franke2016probabilistic}%
\begin{APACrefauthors}%
Franke, M.%
\BCBT {}\ \BBA {} J{\"a}ger, G.%
\end{APACrefauthors}%
\unskip\
\newblock
\APACrefYearMonthDay{2016}{}{}.
\newblock
{\BBOQ}\APACrefatitle {Probabilistic pragmatics, or why Bayes’ rule is
  probably important for pragmatics} {Probabilistic pragmatics, or why bayes’
  rule is probably important for pragmatics}.{\BBCQ}
\newblock
\APACjournalVolNumPages{Zeitschrift f{\"u}r sprachwissenschaft}{35}{1}{3--44}.
\PrintBackRefs{\CurrentBib}

\bibitem [\protect \citeauthoryear {%
Gal%
\ \BBA {} Ghahramani%
}{%
Gal%
\ \BBA {} Ghahramani%
}{%
{\protect \APACyear {2015}}%
}]{%
gal2015dropout}
\APACinsertmetastar {%
gal2015dropout}%
\begin{APACrefauthors}%
Gal, Y.%
\BCBT {}\ \BBA {} Ghahramani, Z.%
\end{APACrefauthors}%
\unskip\
\newblock
\APACrefYearMonthDay{2015}{}{}.
\newblock
{\BBOQ}\APACrefatitle {Dropout as a Bayesian approximation: Insights and
  applications} {Dropout as a bayesian approximation: Insights and
  applications}.{\BBCQ}
\newblock
\BIn{} \APACrefbtitle {Deep Learning Workshop, ICML} {Deep learning workshop,
  icml}\ (\BVOL~1, \BPG~2).
\PrintBackRefs{\CurrentBib}

\bibitem [\protect \citeauthoryear {%
Galantucci%
}{%
Galantucci%
}{%
{\protect \APACyear {2005}}%
}]{%
Galantucci:2005uh}
\APACinsertmetastar {%
Galantucci:2005uh}%
\begin{APACrefauthors}%
Galantucci, B.%
\end{APACrefauthors}%
\unskip\
\newblock
\APACrefYearMonthDay{2005}{}{}.
\newblock
{\BBOQ}\APACrefatitle {{An experimental study of the emergence of human
  communication systems}} {{An experimental study of the emergence of human
  communication systems}}.{\BBCQ}
\newblock
\APACjournalVolNumPages{Cognitive Science}{29}{5}{737--767}.
\PrintBackRefs{\CurrentBib}

\bibitem [\protect \citeauthoryear {%
Ganin%
, Kulkarni%
, Babuschkin%
, Eslami%
\BCBL {}\ \BBA {} Vinyals%
}{%
Ganin%
\ \protect \BOthers {.}}{%
{\protect \APACyear {2018}}%
}]{%
ganin2018synthesizing}
\APACinsertmetastar {%
ganin2018synthesizing}%
\begin{APACrefauthors}%
Ganin, Y.%
, Kulkarni, T.%
, Babuschkin, I.%
, Eslami, S.%
\BCBL {}\ \BBA {} Vinyals, O.%
\end{APACrefauthors}%
\unskip\
\newblock
\APACrefYearMonthDay{2018}{}{}.
\newblock
{\BBOQ}\APACrefatitle {Synthesizing Programs for Images using Reinforced
  Adversarial Learning} {Synthesizing programs for images using reinforced
  adversarial learning}.{\BBCQ}
\newblock
\APACjournalVolNumPages{arXiv preprint arXiv:1804.01118}{}{}{}.
\PrintBackRefs{\CurrentBib}

\bibitem [\protect \citeauthoryear {%
Garrod%
, Fay%
, Lee%
, Oberlander%
\BCBL {}\ \BBA {} MacLeod%
}{%
Garrod%
\ \protect \BOthers {.}}{%
{\protect \APACyear {2007}}%
}]{%
Garrod:2007wk}
\APACinsertmetastar {%
Garrod:2007wk}%
\begin{APACrefauthors}%
Garrod, S.%
, Fay, N.%
, Lee, J.%
, Oberlander, J.%
\BCBL {}\ \BBA {} MacLeod, T.%
\end{APACrefauthors}%
\unskip\
\newblock
\APACrefYearMonthDay{2007}{}{}.
\newblock
{\BBOQ}\APACrefatitle {{Foundations of representation: where might graphical
  symbol systems come from?}} {{Foundations of representation: where might
  graphical symbol systems come from?}}{\BBCQ}
\newblock
\APACjournalVolNumPages{Cognitive science}{31}{6}{961--987}.
\PrintBackRefs{\CurrentBib}

\bibitem [\protect \citeauthoryear {%
Garrod%
, Fay%
, Rogers%
, Walker%
\BCBL {}\ \BBA {} Swoboda%
}{%
Garrod%
\ \protect \BOthers {.}}{%
{\protect \APACyear {2010}}%
}]{%
garrod2010can}
\APACinsertmetastar {%
garrod2010can}%
\begin{APACrefauthors}%
Garrod, S.%
, Fay, N.%
, Rogers, S.%
, Walker, B.%
\BCBL {}\ \BBA {} Swoboda, N.%
\end{APACrefauthors}%
\unskip\
\newblock
\APACrefYearMonthDay{2010}{}{}.
\newblock
{\BBOQ}\APACrefatitle {Can iterated learning explain the emergence of graphical
  symbols?} {Can iterated learning explain the emergence of graphical
  symbols?}{\BBCQ}
\newblock
\APACjournalVolNumPages{Interaction Studies}{11}{1}{33--50}.
\PrintBackRefs{\CurrentBib}

\bibitem [\protect \citeauthoryear {%
Gibson%
}{%
Gibson%
}{%
{\protect \APACyear {2014}}%
}]{%
gibson2014ecological}
\APACinsertmetastar {%
gibson2014ecological}%
\begin{APACrefauthors}%
Gibson, J\BPBI J.%
\end{APACrefauthors}%
\unskip\
\newblock
\APACrefYear{2014}.
\newblock
\APACrefbtitle {The ecological approach to visual perception: classic edition}
  {The ecological approach to visual perception: classic edition}.
\newblock
\APACaddressPublisher{}{Psychology Press}.
\PrintBackRefs{\CurrentBib}

\bibitem [\protect \citeauthoryear {%
Goldin-Meadow%
\ \BBA {} Feldman%
}{%
Goldin-Meadow%
\ \BBA {} Feldman%
}{%
{\protect \APACyear {1977}}%
}]{%
goldin1977development}
\APACinsertmetastar {%
goldin1977development}%
\begin{APACrefauthors}%
Goldin-Meadow, S.%
\BCBT {}\ \BBA {} Feldman, H.%
\end{APACrefauthors}%
\unskip\
\newblock
\APACrefYearMonthDay{1977}{}{}.
\newblock
{\BBOQ}\APACrefatitle {The development of language-like communication without a
  language model} {The development of language-like communication without a
  language model}.{\BBCQ}
\newblock
\APACjournalVolNumPages{Science}{197}{4301}{401--403}.
\PrintBackRefs{\CurrentBib}

\bibitem [\protect \citeauthoryear {%
Gombrich%
}{%
Gombrich%
}{%
{\protect \APACyear {1969}}%
}]{%
gombrich1969art}
\APACinsertmetastar {%
gombrich1969art}%
\begin{APACrefauthors}%
Gombrich, E.%
\end{APACrefauthors}%
\unskip\
\newblock
\APACrefYear{1969}.
\newblock
\APACrefbtitle {Art and Illusion: A Study in the Psychology of Pictorial
  Representation} {Art and illusion: A study in the psychology of pictorial
  representation}.
\newblock
\APACaddressPublisher{}{Princeton University Press}.
\PrintBackRefs{\CurrentBib}

\bibitem [\protect \citeauthoryear {%
Gombrich%
}{%
Gombrich%
}{%
{\protect \APACyear {1989}}%
}]{%
gombrich1989story}
\APACinsertmetastar {%
gombrich1989story}%
\begin{APACrefauthors}%
Gombrich, E.%
\end{APACrefauthors}%
\unskip\
\newblock
\APACrefYear{1989}.
\newblock
\APACrefbtitle {The story of art} {The story of art}.
\newblock
\APACaddressPublisher{}{Phaidon Press, Ltd.}
\PrintBackRefs{\CurrentBib}

\bibitem [\protect \citeauthoryear {%
N.~Goodman%
}{%
N.~Goodman%
}{%
{\protect \APACyear {1976}}%
}]{%
goodman1976languages}
\APACinsertmetastar {%
goodman1976languages}%
\begin{APACrefauthors}%
Goodman, N.%
\end{APACrefauthors}%
\unskip\
\newblock
\APACrefYear{1976}.
\newblock
\APACrefbtitle {Languages of art: An approach to a theory of symbols}
  {Languages of art: An approach to a theory of symbols}.
\newblock
\APACaddressPublisher{}{Hackett publishing}.
\PrintBackRefs{\CurrentBib}

\bibitem [\protect \citeauthoryear {%
N\BPBI D.~Goodman%
\ \BBA {} Frank%
}{%
N\BPBI D.~Goodman%
\ \BBA {} Frank%
}{%
{\protect \APACyear {2016}}%
}]{%
goodman2016pragmatic}
\APACinsertmetastar {%
goodman2016pragmatic}%
\begin{APACrefauthors}%
Goodman, N\BPBI D.%
\BCBT {}\ \BBA {} Frank, M\BPBI C.%
\end{APACrefauthors}%
\unskip\
\newblock
\APACrefYearMonthDay{2016}{}{}.
\newblock
{\BBOQ}\APACrefatitle {Pragmatic language interpretation as probabilistic
  inference} {Pragmatic language interpretation as probabilistic
  inference}.{\BBCQ}
\newblock
\APACjournalVolNumPages{Trends in Cognitive Sciences}{20}{11}{818--829}.
\PrintBackRefs{\CurrentBib}

\bibitem [\protect \citeauthoryear {%
N\BPBI D.~Goodman%
\ \BBA {} Stuhlm{\"u}ller%
}{%
N\BPBI D.~Goodman%
\ \BBA {} Stuhlm{\"u}ller%
}{%
{\protect \APACyear {2013}}%
}]{%
goodman2013knowledge}
\APACinsertmetastar {%
goodman2013knowledge}%
\begin{APACrefauthors}%
Goodman, N\BPBI D.%
\BCBT {}\ \BBA {} Stuhlm{\"u}ller, A.%
\end{APACrefauthors}%
\unskip\
\newblock
\APACrefYearMonthDay{2013}{}{}.
\newblock
{\BBOQ}\APACrefatitle {Knowledge and implicature: Modeling language
  understanding as social cognition} {Knowledge and implicature: Modeling
  language understanding as social cognition}.{\BBCQ}
\newblock
\APACjournalVolNumPages{Topics in cognitive science}{5}{1}{173--184}.
\PrintBackRefs{\CurrentBib}

\bibitem [\protect \citeauthoryear {%
N\BPBI D.~Goodman%
\ \BBA {} Stuhlm{\"u}ller%
}{%
N\BPBI D.~Goodman%
\ \BBA {} Stuhlm{\"u}ller%
}{%
{\protect \APACyear {2014}}%
}]{%
goodman2014design}
\APACinsertmetastar {%
goodman2014design}%
\begin{APACrefauthors}%
Goodman, N\BPBI D.%
\BCBT {}\ \BBA {} Stuhlm{\"u}ller, A.%
\end{APACrefauthors}%
\unskip\
\newblock
\APACrefYearMonthDay{2014}{}{}.
\newblock
\APACrefbtitle {The design and implementation of probabilistic programming
  languages.} {The design and implementation of probabilistic programming
  languages.}
\newblock
\APACaddressPublisher{}{Retrieved 2015/1/16, from http://dippl. org}.
\PrintBackRefs{\CurrentBib}

\bibitem [\protect \citeauthoryear {%
Grice%
, Cole%
\BCBL {}\ \BBA {} Morgan%
}{%
Grice%
\ \protect \BOthers {.}}{%
{\protect \APACyear {1975}}%
}]{%
grice1975syntax}
\APACinsertmetastar {%
grice1975syntax}%
\begin{APACrefauthors}%
Grice, H\BPBI P.%
, Cole, P.%
\BCBL {}\ \BBA {} Morgan, J\BPBI L.%
\end{APACrefauthors}%
\unskip\
\newblock
\APACrefYearMonthDay{1975}{}{}.
\newblock
\APACrefbtitle {Syntax and semantics.} {Syntax and semantics.}
\PrintBackRefs{\CurrentBib}

\bibitem [\protect \citeauthoryear {%
Ha%
\ \BBA {} Eck%
}{%
Ha%
\ \BBA {} Eck%
}{%
{\protect \APACyear {2017}}%
}]{%
ha2017neural}
\APACinsertmetastar {%
ha2017neural}%
\begin{APACrefauthors}%
Ha, D.%
\BCBT {}\ \BBA {} Eck, D.%
\end{APACrefauthors}%
\unskip\
\newblock
\APACrefYearMonthDay{2017}{}{}.
\newblock
{\BBOQ}\APACrefatitle {A neural representation of sketch drawings} {A neural
  representation of sketch drawings}.{\BBCQ}
\newblock
\APACjournalVolNumPages{arXiv preprint arXiv:1704.03477}{}{}{}.
\PrintBackRefs{\CurrentBib}

\bibitem [\protect \citeauthoryear {%
Hinton%
, Srivastava%
, Krizhevsky%
, Sutskever%
\BCBL {}\ \BBA {} Salakhutdinov%
}{%
Hinton%
\ \protect \BOthers {.}}{%
{\protect \APACyear {2012}}%
}]{%
hinton2012improving}
\APACinsertmetastar {%
hinton2012improving}%
\begin{APACrefauthors}%
Hinton, G\BPBI E.%
, Srivastava, N.%
, Krizhevsky, A.%
, Sutskever, I.%
\BCBL {}\ \BBA {} Salakhutdinov, R\BPBI R.%
\end{APACrefauthors}%
\unskip\
\newblock
\APACrefYearMonthDay{2012}{}{}.
\newblock
{\BBOQ}\APACrefatitle {Improving neural networks by preventing co-adaptation of
  feature detectors} {Improving neural networks by preventing co-adaptation of
  feature detectors}.{\BBCQ}
\newblock
\APACjournalVolNumPages{arXiv preprint arXiv:1207.0580}{}{}{}.
\PrintBackRefs{\CurrentBib}

\bibitem [\protect \citeauthoryear {%
Hochberg%
\ \BBA {} Brooks%
}{%
Hochberg%
\ \BBA {} Brooks%
}{%
{\protect \APACyear {1962}}%
}]{%
hochberg1962pictorial}
\APACinsertmetastar {%
hochberg1962pictorial}%
\begin{APACrefauthors}%
Hochberg, J.%
\BCBT {}\ \BBA {} Brooks, V.%
\end{APACrefauthors}%
\unskip\
\newblock
\APACrefYearMonthDay{1962}{}{}.
\newblock
{\BBOQ}\APACrefatitle {Pictorial recognition as an unlearned ability: A study
  of one child's performance} {Pictorial recognition as an unlearned ability: A
  study of one child's performance}.{\BBCQ}
\newblock
\APACjournalVolNumPages{the american Journal of Psychology}{}{}{624--628}.
\PrintBackRefs{\CurrentBib}

\bibitem [\protect \citeauthoryear {%
Hoffmann%
\ \protect \BOthers {.}}{%
Hoffmann%
\ \protect \BOthers {.}}{%
{\protect \APACyear {2018}}%
}]{%
hoffmann2018u}
\APACinsertmetastar {%
hoffmann2018u}%
\begin{APACrefauthors}%
Hoffmann, D.%
, Standish, C.%
, Garc{\'\i}a-Diez, M.%
, Pettitt, P.%
, Milton, J.%
, Zilh{\~a}o, J.%
\BDBL {}others%
\end{APACrefauthors}%
\unskip\
\newblock
\APACrefYearMonthDay{2018}{}{}.
\newblock
{\BBOQ}\APACrefatitle {U-Th dating of carbonate crusts reveals Neandertal
  origin of Iberian cave art} {U-th dating of carbonate crusts reveals
  neandertal origin of iberian cave art}.{\BBCQ}
\newblock
\APACjournalVolNumPages{Science}{359}{6378}{912--915}.
\PrintBackRefs{\CurrentBib}

\bibitem [\protect \citeauthoryear {%
Hornik%
}{%
Hornik%
}{%
{\protect \APACyear {1991}}%
}]{%
hornik1991approximation}
\APACinsertmetastar {%
hornik1991approximation}%
\begin{APACrefauthors}%
Hornik, K.%
\end{APACrefauthors}%
\unskip\
\newblock
\APACrefYearMonthDay{1991}{}{}.
\newblock
{\BBOQ}\APACrefatitle {Approximation capabilities of multilayer feedforward
  networks} {Approximation capabilities of multilayer feedforward
  networks}.{\BBCQ}
\newblock
\APACjournalVolNumPages{Neural networks}{4}{2}{251--257}.
\PrintBackRefs{\CurrentBib}

\bibitem [\protect \citeauthoryear {%
Jefferys%
\ \BBA {} Berger%
}{%
Jefferys%
\ \BBA {} Berger%
}{%
{\protect \APACyear {1992}}%
}]{%
jefferys1992ockham}
\APACinsertmetastar {%
jefferys1992ockham}%
\begin{APACrefauthors}%
Jefferys, W\BPBI H.%
\BCBT {}\ \BBA {} Berger, J\BPBI O.%
\end{APACrefauthors}%
\unskip\
\newblock
\APACrefYearMonthDay{1992}{}{}.
\newblock
{\BBOQ}\APACrefatitle {Ockham's razor and Bayesian analysis} {Ockham's razor
  and bayesian analysis}.{\BBCQ}
\newblock
\APACjournalVolNumPages{American Scientist}{80}{1}{64--72}.
\PrintBackRefs{\CurrentBib}

\bibitem [\protect \citeauthoryear {%
Kao%
, Bergen%
\BCBL {}\ \BBA {} Goodman%
}{%
Kao%
\ \protect \BOthers {.}}{%
{\protect \APACyear {2014}}%
}]{%
kao2014formalizing}
\APACinsertmetastar {%
kao2014formalizing}%
\begin{APACrefauthors}%
Kao, J.%
, Bergen, L.%
\BCBL {}\ \BBA {} Goodman, N.%
\end{APACrefauthors}%
\unskip\
\newblock
\APACrefYearMonthDay{2014}{}{}.
\newblock
{\BBOQ}\APACrefatitle {Formalizing the pragmatics of metaphor understanding}
  {Formalizing the pragmatics of metaphor understanding}.{\BBCQ}
\newblock
\BIn{} \APACrefbtitle {Proceedings of the annual meeting of the Cognitive
  Science Society} {Proceedings of the annual meeting of the cognitive science
  society}\ (\BVOL~36).
\PrintBackRefs{\CurrentBib}

\bibitem [\protect \citeauthoryear {%
Kennedy%
\ \BBA {} Ross%
}{%
Kennedy%
\ \BBA {} Ross%
}{%
{\protect \APACyear {1975}}%
}]{%
kennedy1975outline}
\APACinsertmetastar {%
kennedy1975outline}%
\begin{APACrefauthors}%
Kennedy, J\BPBI M.%
\BCBT {}\ \BBA {} Ross, A\BPBI S.%
\end{APACrefauthors}%
\unskip\
\newblock
\APACrefYearMonthDay{1975}{}{}.
\newblock
{\BBOQ}\APACrefatitle {Outline picture perception by the Songe of Papua}
  {Outline picture perception by the songe of papua}.{\BBCQ}
\newblock
\APACjournalVolNumPages{Perception}{4}{4}{391--406}.
\PrintBackRefs{\CurrentBib}

\bibitem [\protect \citeauthoryear {%
Kingma%
\ \BBA {} Ba%
}{%
Kingma%
\ \BBA {} Ba%
}{%
{\protect \APACyear {2014}}%
}]{%
kingma2014adam}
\APACinsertmetastar {%
kingma2014adam}%
\begin{APACrefauthors}%
Kingma, D\BPBI P.%
\BCBT {}\ \BBA {} Ba, J.%
\end{APACrefauthors}%
\unskip\
\newblock
\APACrefYearMonthDay{2014}{}{}.
\newblock
{\BBOQ}\APACrefatitle {Adam: A method for stochastic optimization} {Adam: A
  method for stochastic optimization}.{\BBCQ}
\newblock
\APACjournalVolNumPages{arXiv preprint arXiv:1412.6980}{}{}{}.
\PrintBackRefs{\CurrentBib}

\bibitem [\protect \citeauthoryear {%
Kubilius%
, Bracci%
\BCBL {}\ \BBA {} de Beeck%
}{%
Kubilius%
\ \protect \BOthers {.}}{%
{\protect \APACyear {2016}}%
}]{%
kubilius2016deep}
\APACinsertmetastar {%
kubilius2016deep}%
\begin{APACrefauthors}%
Kubilius, J.%
, Bracci, S.%
\BCBL {}\ \BBA {} de Beeck, H\BPBI P\BPBI O.%
\end{APACrefauthors}%
\unskip\
\newblock
\APACrefYearMonthDay{2016}{}{}.
\newblock
{\BBOQ}\APACrefatitle {Deep neural networks as a computational model for human
  shape sensitivity} {Deep neural networks as a computational model for human
  shape sensitivity}.{\BBCQ}
\newblock
\APACjournalVolNumPages{PLoS computational biology}{12}{4}{e1004896}.
\PrintBackRefs{\CurrentBib}

\bibitem [\protect \citeauthoryear {%
Lake%
, Salakhutdinov%
\BCBL {}\ \BBA {} Tenenbaum%
}{%
Lake%
\ \protect \BOthers {.}}{%
{\protect \APACyear {2015}}%
}]{%
lake2015human}
\APACinsertmetastar {%
lake2015human}%
\begin{APACrefauthors}%
Lake, B\BPBI M.%
, Salakhutdinov, R.%
\BCBL {}\ \BBA {} Tenenbaum, J\BPBI B.%
\end{APACrefauthors}%
\unskip\
\newblock
\APACrefYearMonthDay{2015}{}{}.
\newblock
{\BBOQ}\APACrefatitle {Human-level concept learning through probabilistic
  program induction} {Human-level concept learning through probabilistic
  program induction}.{\BBCQ}
\newblock
\APACjournalVolNumPages{Science}{350}{6266}{1332--1338}.
\PrintBackRefs{\CurrentBib}

\bibitem [\protect \citeauthoryear {%
Lewis%
}{%
Lewis%
}{%
{\protect \APACyear {1969}}%
}]{%
Lewis69_Convention}
\APACinsertmetastar {%
Lewis69_Convention}%
\begin{APACrefauthors}%
Lewis, D.%
\end{APACrefauthors}%
\unskip\
\newblock
\APACrefYear{1969}.
\newblock
\APACrefbtitle {Convention: A philosophical study} {Convention: A philosophical
  study}.
\newblock
\APACaddressPublisher{}{Harvard University Press}.
\PrintBackRefs{\CurrentBib}

\bibitem [\protect \citeauthoryear {%
Monroe%
, Hawkins%
, Goodman%
\BCBL {}\ \BBA {} Potts%
}{%
Monroe%
\ \protect \BOthers {.}}{%
{\protect \APACyear {2017}}%
}]{%
monroe2017colors}
\APACinsertmetastar {%
monroe2017colors}%
\begin{APACrefauthors}%
Monroe, W.%
, Hawkins, R\BPBI X.%
, Goodman, N\BPBI D.%
\BCBL {}\ \BBA {} Potts, C.%
\end{APACrefauthors}%
\unskip\
\newblock
\APACrefYearMonthDay{2017}{}{}.
\newblock
{\BBOQ}\APACrefatitle {Colors in context: A pragmatic neural model for grounded
  language understanding} {Colors in context: A pragmatic neural model for
  grounded language understanding}.{\BBCQ}
\newblock
\APACjournalVolNumPages{arXiv preprint arXiv:1703.10186}{}{}{}.
\PrintBackRefs{\CurrentBib}

\bibitem [\protect \citeauthoryear {%
Mrowca%
\ \protect \BOthers {.}}{%
Mrowca%
\ \protect \BOthers {.}}{%
{\protect \APACyear {2018}}%
}]{%
mrowca2018graph}
\APACinsertmetastar {%
mrowca2018graph}%
\begin{APACrefauthors}%
Mrowca, D.%
, Zhuang, C.%
, Wang, E.%
, Haber, N.%
, Fei-Fei, L.%
, Tenenbaum, J\BPBI B.%
\BCBL {}\ \BBA {} Yamins, D\BPBI L.%
\end{APACrefauthors}%
\unskip\
\newblock
\APACrefYearMonthDay{2018}{}{}.
\newblock
{\BBOQ}\APACrefatitle {Flexible Neural Representation for Physics Prediction}
  {Flexible neural representation for physics prediction}.{\BBCQ}
\newblock
\BIn{} \APACrefbtitle {{Advances in Neural Information Processing Systems}.}
  {{Advances in Neural Information Processing Systems}.}
\PrintBackRefs{\CurrentBib}

\bibitem [\protect \citeauthoryear {%
Ramachandran%
, Zoph%
\BCBL {}\ \BBA {} Le%
}{%
Ramachandran%
\ \protect \BOthers {.}}{%
{\protect \APACyear {2018}}%
}]{%
ramachandran2018searching}
\APACinsertmetastar {%
ramachandran2018searching}%
\begin{APACrefauthors}%
Ramachandran, P.%
, Zoph, B.%
\BCBL {}\ \BBA {} Le, Q\BPBI V.%
\end{APACrefauthors}%
\unskip\
\newblock
\APACrefYearMonthDay{2018}{}{}.
\newblock
{\BBOQ}\APACrefatitle {Searching for activation functions} {Searching for
  activation functions}.{\BBCQ}
\newblock

\PrintBackRefs{\CurrentBib}

\bibitem [\protect \citeauthoryear {%
Sangkloy%
, Burnell%
, Ham%
\BCBL {}\ \BBA {} Hays%
}{%
Sangkloy%
\ \protect \BOthers {.}}{%
{\protect \APACyear {2016}}%
}]{%
sangkloy2016sketchy}
\APACinsertmetastar {%
sangkloy2016sketchy}%
\begin{APACrefauthors}%
Sangkloy, P.%
, Burnell, N.%
, Ham, C.%
\BCBL {}\ \BBA {} Hays, J.%
\end{APACrefauthors}%
\unskip\
\newblock
\APACrefYearMonthDay{2016}{}{}.
\newblock
{\BBOQ}\APACrefatitle {The sketchy database: learning to retrieve badly drawn
  bunnies} {The sketchy database: learning to retrieve badly drawn
  bunnies}.{\BBCQ}
\newblock
\APACjournalVolNumPages{ACM Transactions on Graphics (TOG)}{35}{4}{119}.
\PrintBackRefs{\CurrentBib}

\bibitem [\protect \citeauthoryear {%
Sayim%
}{%
Sayim%
}{%
{\protect \APACyear {2011}}%
}]{%
Sayim:2011bz}
\APACinsertmetastar {%
Sayim:2011bz}%
\begin{APACrefauthors}%
Sayim, B.%
\end{APACrefauthors}%
\unskip\
\newblock
\APACrefYearMonthDay{2011}{{\APACmonth{10}}}{}.
\newblock
{\BBOQ}\APACrefatitle {{What line drawings reveal about the visual brain}}
  {{What line drawings reveal about the visual brain}}.{\BBCQ}
\newblock
\APACjournalVolNumPages{}{}{}{1--4}.
\PrintBackRefs{\CurrentBib}

\bibitem [\protect \citeauthoryear {%
Simonyan%
\ \BBA {} Zisserman%
}{%
Simonyan%
\ \BBA {} Zisserman%
}{%
{\protect \APACyear {2014}}%
}]{%
simonyan2014very}
\APACinsertmetastar {%
simonyan2014very}%
\begin{APACrefauthors}%
Simonyan, K.%
\BCBT {}\ \BBA {} Zisserman, A.%
\end{APACrefauthors}%
\unskip\
\newblock
\APACrefYearMonthDay{2014}{}{}.
\newblock
{\BBOQ}\APACrefatitle {Very deep convolutional networks for large-scale image
  recognition} {Very deep convolutional networks for large-scale image
  recognition}.{\BBCQ}
\newblock
\APACjournalVolNumPages{arXiv preprint arXiv:1409.1556}{}{}{}.
\PrintBackRefs{\CurrentBib}

\bibitem [\protect \citeauthoryear {%
Song%
, Yu%
, Song%
, Xiang%
\BCBL {}\ \BBA {} Hospedales%
}{%
Song%
\ \protect \BOthers {.}}{%
{\protect \APACyear {2017}}%
}]{%
song2017deep}
\APACinsertmetastar {%
song2017deep}%
\begin{APACrefauthors}%
Song, J.%
, Yu, Q.%
, Song, Y\BHBI Z.%
, Xiang, T.%
\BCBL {}\ \BBA {} Hospedales, T\BPBI M.%
\end{APACrefauthors}%
\unskip\
\newblock
\APACrefYearMonthDay{2017}{}{}.
\newblock
{\BBOQ}\APACrefatitle {Deep Spatial-Semantic Attention for Fine-Grained
  Sketch-Based Image Retrieval.} {Deep spatial-semantic attention for
  fine-grained sketch-based image retrieval.}{\BBCQ}
\newblock
\BIn{} \APACrefbtitle {ICCV} {Iccv}\ (\BPGS\ 5552--5561).
\PrintBackRefs{\CurrentBib}

\bibitem [\protect \citeauthoryear {%
Tanaka%
}{%
Tanaka%
}{%
{\protect \APACyear {2007}}%
}]{%
tanaka2007recognition}
\APACinsertmetastar {%
tanaka2007recognition}%
\begin{APACrefauthors}%
Tanaka, M.%
\end{APACrefauthors}%
\unskip\
\newblock
\APACrefYearMonthDay{2007}{}{}.
\newblock
{\BBOQ}\APACrefatitle {Recognition of pictorial representations by chimpanzees
  (Pan troglodytes)} {Recognition of pictorial representations by chimpanzees
  (pan troglodytes)}.{\BBCQ}
\newblock
\APACjournalVolNumPages{Animal cognition}{10}{2}{169--179}.
\PrintBackRefs{\CurrentBib}

\bibitem [\protect \citeauthoryear {%
Theisen%
, Oberlander%
\BCBL {}\ \BBA {} Kirby%
}{%
Theisen%
\ \protect \BOthers {.}}{%
{\protect \APACyear {2010}}%
}]{%
theisen2010systematicity}
\APACinsertmetastar {%
theisen2010systematicity}%
\begin{APACrefauthors}%
Theisen, C\BPBI A.%
, Oberlander, J.%
\BCBL {}\ \BBA {} Kirby, S.%
\end{APACrefauthors}%
\unskip\
\newblock
\APACrefYearMonthDay{2010}{}{}.
\newblock
{\BBOQ}\APACrefatitle {Systematicity and arbitrariness in novel communication
  systems} {Systematicity and arbitrariness in novel communication
  systems}.{\BBCQ}
\newblock
\APACjournalVolNumPages{Interaction Studies}{11}{1}{14--32}.
\PrintBackRefs{\CurrentBib}

\bibitem [\protect \citeauthoryear {%
Tomasello%
}{%
Tomasello%
}{%
{\protect \APACyear {2009}}%
}]{%
tomasello2009cultural}
\APACinsertmetastar {%
tomasello2009cultural}%
\begin{APACrefauthors}%
Tomasello, M.%
\end{APACrefauthors}%
\unskip\
\newblock
\APACrefYear{2009}.
\newblock
\APACrefbtitle {The cultural origins of human cognition} {The cultural origins
  of human cognition}.
\newblock
\APACaddressPublisher{}{Harvard university press}.
\PrintBackRefs{\CurrentBib}

\bibitem [\protect \citeauthoryear {%
Verhoef%
, Kirby%
\BCBL {}\ \BBA {} De~Boer%
}{%
Verhoef%
\ \protect \BOthers {.}}{%
{\protect \APACyear {2014}}%
}]{%
verhoef2014emergence}
\APACinsertmetastar {%
verhoef2014emergence}%
\begin{APACrefauthors}%
Verhoef, T.%
, Kirby, S.%
\BCBL {}\ \BBA {} De~Boer, B.%
\end{APACrefauthors}%
\unskip\
\newblock
\APACrefYearMonthDay{2014}{}{}.
\newblock
{\BBOQ}\APACrefatitle {Emergence of combinatorial structure and economy through
  iterated learning with continuous acoustic signals} {Emergence of
  combinatorial structure and economy through iterated learning with continuous
  acoustic signals}.{\BBCQ}
\newblock
\APACjournalVolNumPages{Journal of Phonetics}{43}{}{57--68}.
\PrintBackRefs{\CurrentBib}

\bibitem [\protect \citeauthoryear {%
Wagenmakers%
, Lodewyckx%
, Kuriyal%
\BCBL {}\ \BBA {} Grasman%
}{%
Wagenmakers%
\ \protect \BOthers {.}}{%
{\protect \APACyear {2010}}%
}]{%
wagenmakers2010bayesian}
\APACinsertmetastar {%
wagenmakers2010bayesian}%
\begin{APACrefauthors}%
Wagenmakers, E\BHBI J.%
, Lodewyckx, T.%
, Kuriyal, H.%
\BCBL {}\ \BBA {} Grasman, R.%
\end{APACrefauthors}%
\unskip\
\newblock
\APACrefYearMonthDay{2010}{}{}.
\newblock
{\BBOQ}\APACrefatitle {Bayesian hypothesis testing for psychologists: A
  tutorial on the Savage--Dickey method} {Bayesian hypothesis testing for
  psychologists: A tutorial on the savage--dickey method}.{\BBCQ}
\newblock
\APACjournalVolNumPages{Cognitive psychology}{60}{3}{158--189}.
\PrintBackRefs{\CurrentBib}

\bibitem [\protect \citeauthoryear {%
Wagenmakers%
\ \protect \BOthers {.}}{%
Wagenmakers%
\ \protect \BOthers {.}}{%
{\protect \APACyear {2018}}%
}]{%
wagenmakers2018bayesian}
\APACinsertmetastar {%
wagenmakers2018bayesian}%
\begin{APACrefauthors}%
Wagenmakers, E\BHBI J.%
, Marsman, M.%
, Jamil, T.%
, Ly, A.%
, Verhagen, J.%
, Love, J.%
\BDBL {}others%
\end{APACrefauthors}%
\unskip\
\newblock
\APACrefYearMonthDay{2018}{}{}.
\newblock
{\BBOQ}\APACrefatitle {Bayesian inference for psychology. Part I: Theoretical
  advantages and practical ramifications} {Bayesian inference for psychology.
  part i: Theoretical advantages and practical ramifications}.{\BBCQ}
\newblock
\APACjournalVolNumPages{Psychonomic bulletin \& review}{25}{1}{35--57}.
\PrintBackRefs{\CurrentBib}

\bibitem [\protect \citeauthoryear {%
Wilson%
\ \BBA {} Sperber%
}{%
Wilson%
\ \BBA {} Sperber%
}{%
{\protect \APACyear {1986}}%
}]{%
wilson1986relevance}
\APACinsertmetastar {%
wilson1986relevance}%
\begin{APACrefauthors}%
Wilson, D.%
\BCBT {}\ \BBA {} Sperber, D.%
\end{APACrefauthors}%
\unskip\
\newblock
\APACrefYear{1986}.
\newblock
\APACrefbtitle {Relevance: Communication and cognition} {Relevance:
  Communication and cognition}.
\newblock
\APACaddressPublisher{}{Mass.}
\PrintBackRefs{\CurrentBib}

\bibitem [\protect \citeauthoryear {%
Wittgenstein%
}{%
Wittgenstein%
}{%
{\protect \APACyear {1953}}%
}]{%
wittgenstein1953philosophical}
\APACinsertmetastar {%
wittgenstein1953philosophical}%
\begin{APACrefauthors}%
Wittgenstein, L.%
\end{APACrefauthors}%
\unskip\
\newblock
\APACrefYear{1953}.
\newblock
\APACrefbtitle {Philosophical investigations} {Philosophical investigations}.
\newblock
\APACaddressPublisher{}{Macmillan}.
\PrintBackRefs{\CurrentBib}

\bibitem [\protect \citeauthoryear {%
Xu%
\ \protect \BOthers {.}}{%
Xu%
\ \protect \BOthers {.}}{%
{\protect \APACyear {2015}}%
}]{%
xu2015show}
\APACinsertmetastar {%
xu2015show}%
\begin{APACrefauthors}%
Xu, K.%
, Ba, J.%
, Kiros, R.%
, Cho, K.%
, Courville, A.%
, Salakhudinov, R.%
\BDBL {}Bengio, Y.%
\end{APACrefauthors}%
\unskip\
\newblock
\APACrefYearMonthDay{2015}{}{}.
\newblock
{\BBOQ}\APACrefatitle {Show, attend and tell: Neural image caption generation
  with visual attention} {Show, attend and tell: Neural image caption
  generation with visual attention}.{\BBCQ}
\newblock
\BIn{} \APACrefbtitle {International conference on machine learning}
  {International conference on machine learning}\ (\BPGS\ 2048--2057).
\PrintBackRefs{\CurrentBib}

\bibitem [\protect \citeauthoryear {%
Yamins%
\ \protect \BOthers {.}}{%
Yamins%
\ \protect \BOthers {.}}{%
{\protect \APACyear {2014}}%
}]{%
yamins2014performance}
\APACinsertmetastar {%
yamins2014performance}%
\begin{APACrefauthors}%
Yamins, D\BPBI L.%
, Hong, H.%
, Cadieu, C\BPBI F.%
, Solomon, E\BPBI A.%
, Seibert, D.%
\BCBL {}\ \BBA {} DiCarlo, J\BPBI J.%
\end{APACrefauthors}%
\unskip\
\newblock
\APACrefYearMonthDay{2014}{}{}.
\newblock
{\BBOQ}\APACrefatitle {Performance-optimized hierarchical models predict neural
  responses in higher visual cortex} {Performance-optimized hierarchical models
  predict neural responses in higher visual cortex}.{\BBCQ}
\newblock
\APACjournalVolNumPages{Proceedings of the National Academy of
  Sciences}{111}{23}{8619--8624}.
\PrintBackRefs{\CurrentBib}

\bibitem [\protect \citeauthoryear {%
Yu%
\ \protect \BOthers {.}}{%
Yu%
\ \protect \BOthers {.}}{%
{\protect \APACyear {2017}}%
}]{%
yu2017sketch}
\APACinsertmetastar {%
yu2017sketch}%
\begin{APACrefauthors}%
Yu, Q.%
, Yang, Y.%
, Liu, F.%
, Song, Y\BHBI Z.%
, Xiang, T.%
\BCBL {}\ \BBA {} Hospedales, T\BPBI M.%
\end{APACrefauthors}%
\unskip\
\newblock
\APACrefYearMonthDay{2017}{}{}.
\newblock
{\BBOQ}\APACrefatitle {Sketch-a-net: A deep neural network that beats humans}
  {Sketch-a-net: A deep neural network that beats humans}.{\BBCQ}
\newblock
\APACjournalVolNumPages{International journal of computer
  vision}{122}{3}{411--425}.
\PrintBackRefs{\CurrentBib}

\bibitem [\protect \citeauthoryear {%
Zipf%
}{%
Zipf%
}{%
{\protect \APACyear {1936}}%
}]{%
zipf1936psycho}
\APACinsertmetastar {%
zipf1936psycho}%
\begin{APACrefauthors}%
Zipf, G\BPBI K.%
\end{APACrefauthors}%
\unskip\
\newblock
\APACrefYear{1936}.
\newblock
\APACrefbtitle {The psycho-biology of language: An introduction to dynamic
  philology} {The psycho-biology of language: An introduction to dynamic
  philology}.
\newblock
\APACaddressPublisher{}{Routledge}.
\PrintBackRefs{\CurrentBib}

\end{thebibliography}

\end{document}